\documentclass[%
 reprint,
superscriptaddress,
 amsmath,amssymb,
 aps,
prx
]{revtex4-2}

\usepackage[dvipsnames]{xcolor}

\newcommand{\rb}[1]{\left(#1\right)}

 \newcommand{\myeq}[1] {
Eq.~\ref{eq:#1}}

 \newcommand{\myfig}[1] {
Fig.~\ref{fig:#1}}

 \newcommand{\nvec} {
\mathbf{n}}

 \newcommand{\zvec} {
\mathbf{z}}

 \newcommand{\uvec} {
\mathbf{u}}

 \newcommand{\rmat} {
\mathbf{R}}

 \newcommand{\evec} {
\mathbf{e}}

\usepackage{graphicx} 
\usepackage[utf8]{inputenc}
\usepackage{amsmath}
\usepackage{braket}
\usepackage{bm}
\usepackage{tikz}
\usepackage{float}
\usepackage{hyperref}
\usepackage{amssymb}
\usepackage[normalem]{ulem}

\date{September 2024}

\begin{abstract}
Origami principles are used to create strong, lightweight structures with complex mechanical response.
However, identifying the fundamental physical principles that determine a sheet's behavior remains a challenge.
We introduce a new analytic theory in which commonly studied origami sheets fall into distinct topological classes that predict sharply varying mechanical behavior, including effective stiffness and smoothness of mechanical response under external loads.
Origami sheets with negative Poisson's ratios, such as the Miura ori, have conventional, smooth mechanical response amenable to continuum-based approaches.
In contrast, positive Poisson's ratio, as in the Eggbox ori, generates a topological transition to lines of doubly degenerate zero modes that lead to dramatically softer structures with uneven, complex patterns of spatial response.
These patterns interact in complicated ways with origami boundary conditions and source terms, leading to rich physical phenomena in experimentally accessible systems.
This approach highlights topological mechanics, with deep connections to topologically protected quantum-mechanical systems, as a design principle for controlling the mechanical response of thin, complex sheets.
\end{abstract}
\begin{document}
\author{Yanxin Feng}
\affiliation{School of Physics, Georgia Institute of Technology, Atlanta, GA 30332, USA}
\author{Andrew Wu}
\affiliation{School of Physics, Georgia Institute of Technology, Atlanta, GA 30332, USA}
\author{James McInerney}
\affiliation{Department of Physics, University of Michigan, Ann Arbor, MI 48109, USA}
\author{Siddhartha Sarkar}
\affiliation{Department of Physics, University of Michigan, Ann Arbor, MI 48109, USA}
\author{Xiaoming Mao}
\affiliation{Department of Physics, University of Michigan, Ann Arbor, MI 48109, USA}
\author{D. Zeb Rocklin}
\affiliation{School of Physics, Georgia Institute of Technology, Atlanta, GA 30332, USA}


\title{Novel mechanical response of parallelogram-face origami governed by topological characteristics}

\maketitle

\section{Introduction}

Origami sheets are two-dimensional structures constructed by generating patterns of creases
along which they can fold at low energy cost. 
Originally developed only for their recreational and aesthetic values, these sheets   have intrigued mathematicians~\cite{Hull2005}, computer scientists~\cite{lang1996computational,demaine2007geometric}, and physicists~\cite{doi:10.1073/pnas.2005089117, doi:10.1073/pnas.2202777119}, as well as engineers who seek novel metamaterials~\cite{article, Lv2014} or to design robots~\cite{doi:10.1126/scirobotics.aan1544, jin2023worm, Yi2022}. The geometry of  origami sheets, as defined by their crease patterns, can result in dramatically different responses when external loads are applied. 
The large space of possible crease patterns allow origami sheets to exhibit exotic mechanical properties beyond those of the original sheets of paper,  such as tunable Poisson's ratio~\cite{MISSERONI2022101685}, multi-directional auxeticity~\cite{WANG2020100715}, programmability~\cite{doi:10.1073/pnas.0914069107} and deployability~\cite{WANG2022113934}, making origami-inspired structures widely applicable in different fields.

Despite these broad applications, predicting and controlling the nonuniform deformations of origami sheets under general loading remains challenging.
To address this,
we take an approach which first became prevalent in analyzing quantum mechanical systems and later was applied to analogous classical mechanical systems---the topological classification. In quantum mechanics, since the discovery of special quantum states~\cite{PhysRevLett.45.494} that are topologically distinct and lead to quantization associated with certain topological invariants~\cite{PhysRevB.23.5632}, physicists have been engaged in looking for systems with novel topological signatures as well as classifying them accordingly. The most well-known classification, the ten-fold way, was established by Altland and Zirnbauer~\cite{Zirnbauer_1996,PhysRevB.55.1142} and there are other variants based on the this, such as the three-fold way introduced for non-Hermitian matrices~\cite{roychowdhury2018classification}. This approach allows physicists to better understand phenomena like the quantum Hall effect and to infer important properties such as the band structure in different quantum systems, including topological insulators~\cite{moore2010birth}. In classical mechanical systems, researchers strive to explore novel mechanical properties indicated by similar topological invariants. For instance, the quantum Su–Schrieffer–Heeger (SSH) model~\cite{su1980soliton} has its classical mechanical counterpart introduced by Kane and Lubensky via a 1-D isostatic lattice~\cite{Kane2014}. ``Topological insulators" were realized in various mechanical systems as well~\cite{susstrunk2015observation, mitchell2018amorphous, wang2015topological}. These works show how edge states are controlled by the topological invariants in the bulk (the bulk-boundary correspondence) and show how a topological approach is necessary to understand the behavior of such systems. In the field of origami, topology can also play a vital role in determining zero-energy deformations~\cite{doi:10.1073/pnas.2005089117} or localized deformation~\cite{PhysRevLett.116.135501} and provide new insights into the analysis of its mechanical properties~\cite{miyazawa2022topological,li2024topological}. While generic 2-D periodic origami sheets are shown to have no topological polarization~\cite{PhysRevLett.116.135501}, we identify a topological invariant, known as the Pfaffian~\cite{cayley1894collected}, in the special case where the unit cell consists of parallelogram panels. This topological invariant has not previously been found in mechanical systems but underlies the Quantum Spin Hall Effect~\cite{kane2005z}.

In this paper, we examine a particular class of origami, one with parallelogram faces, with a focus on the Morph patterns~\cite{PhysRevLett.122.155501}. We find that this geometry induces additional symmetries that lead to novel topological invariants given by the Pfaffian. We find that auxetic origami such as the Miura~\cite{miura2009science, schenk2013geometry} 
is topologically trivial, but that non-auxetic origami such as the Eggbox~\cite{schenk2011origami, nassar2017curvature} has a nontrivial invariant which leads to lines of doubly degenerate zero modes stretching across the Brillouin zone, which strongly modify the origami's mechanical response with modes activated at finite wavelength.

This paper is organized as follows. In Sec.~\ref{sec:main} we introduce our system and characterize its topological properties using a novel formalism based on the origami sheet's symmetries. In Sec.~\ref{sec:response} we consider how interplay between origami stretching and bending informs the response of the system. In Sec.~\ref{sec:numerics} we simulate origami sheets from distinct topological classes and characterize their sharply distinct responses. In Sec.~\ref{sec:discussion} we discuss the implications of our work for the field.

\begin{widetext}

    \begin{figure}[H]
    \centering
    \includegraphics[width=\textwidth]{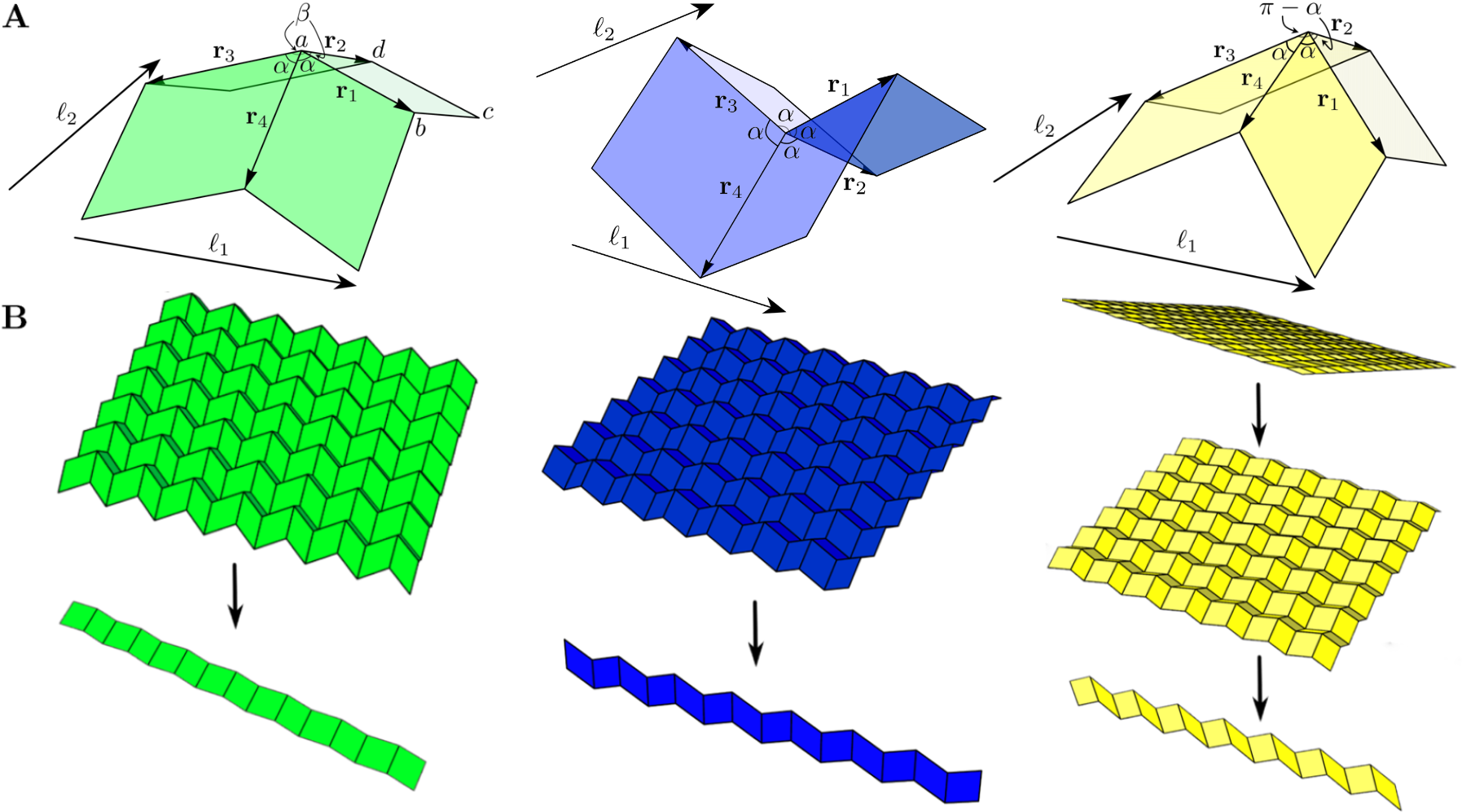}
    \caption{ (A) The geometry of an origami unit cell of the Morph structure, consisting of four parallelogram faces with sector angles $\alpha, \alpha, \beta, \beta$ at a single vertex. Here, $\mathbf{r}_i$ are edge vectors around the vertex $a$ and $\ell_1,\ell_2$ are the two lattice vectors. The generic case (green, left) is known as the Morph, the symmetric case (blue, center) is known as the Eggbox and has four equal angles $\alpha$ and the developable case (yellow, right) is known as the Miura and has $\beta = \pi - \alpha$.
    (B) Folding mechanisms of the different types of origami. A generic Morph pattern is always flat-foldable (meaning that it can be rigidly folded into a flat state), but only the Miura (the yellow one) is developable, meaning that it corresponds to a crease pattern applied to a flat sheet of paper. 
}
    \label{fig:figure1}
\end{figure}

\end{widetext}
\section{PFAFFIAN AS TOPOLOGICAL CHARGE GOVERNS MECHANICAL RESPONSE}
\label{sec:main}

\subsection{Geometry of parallelogram-based origami and description of deformation via folding amplitudes}

We consider origami sheets consisting of repeating unit cells indexed by $\nvec =(n_1,n_2)$, each of which contains four parallelogram faces, as in~\cite{doi:10.1073/pnas.2202777119}. The geometry of these faces is set by their edge lengths and sector angles as in\myfig{figure1} (A). Once the geometry of the faces are fixed, the spatial embedding of the sheet is determined by the dihedral angles of adjoining faces. We restrict ourselves to systems with a common edge length and at most two parallelogram geometries set by their sector angles $\alpha,\beta$, the class of Morph origami~\cite{PhysRevLett.122.155501}. As special cases, we also consider $\beta=\alpha$, the Eggbox crease pattern~\cite{schenk2012folded, filipov2015origami} and $\beta = \pi - \alpha$, the Miura fold~\cite{miura2009science,wei2013geometric}. Examples of such origami sheets as well as their unit cells are shown in\myfig{figure1}. This range of systems suffices to realize the range of topological classes and types of mechanical response on which we focus.
Each vertex of one of these parallelogram-based origami sheets possesses a 1-D family of nonlinear isometries, foldings of the adjoining edges that do not stretch the adjoining faces. The geometry of different vertices is compatible so that each sheet possesses a rigid, uniform folding motion, as shown in\myfig{figure1} (B). For origami sheets made out of conventional materials such as paperboard or metal, the energy cost of stretching the faces of the sheets is much higher than that of bending or folding the sheets. Hence, those isometries are low-energy modes favored by origami sheets, and in the ideal case where we neglect the energy associated with folding and bending, they become the zero-energy modes. 
We thus refer to these isometries as ``zero modes'' despite the inclusion in our numerical modeling of weak bending stiffness that converts the actual physical response to low-energy ``soft modes''.
We consider a linearized but non-uniform version of these modes, in which the degree to which a vertex is folded along this manifold of isometries is given by its folding amplitude $\mathcal{V}$. In a unit cell there are four vertices, labeled $a,b,c,d$ as shown in\myfig{figure1} (A). We may refer to a potential isometric deformation of the sheet by a set of  folding amplitudes at each vertex in each cell, denoted $\mathcal{V}^a(\mathbf{n})$.
Folding two vertices with a shared edge by different amounts implies that the dihedral angle varies along the length of the edge~\cite{doi:10.1073/pnas.2202777119}, and therefore that the adjoining faces are bending (a non-rigid isometry). Conditions for compatible face bending are derived in the Supplementary Materials.

We are particularly interested in $z$-periodic modes~\cite{doi:10.1073/pnas.2005089117, Rocklin2020}, those of the form
\begin{align}
    \mathcal{V}^{a}(\mathbf{n})\equiv\mathcal{V}^a(\mathbf{0})\mathbf{z}^{\mathbf{n}}\equiv\mathcal{V}^a z_1^{n_1}z_2^{n_2},
\end{align}
which extends  Bloch's theorem~\cite{griffiths2018introduction} by allowing the wavevectors to be not purely real so as to permit  edge modes localized at the boundaries of finite sheets. Bulk modes are the special case of $\mathbf{z}=e^{i\mathbf{q}}$ with  $\mathbf{q}$ a purely real wavevector in the Brillouin zone. 

The aforementioned (linearized) global isometries correspond to modes
$z_1=z_2=1$. For a non-uniform global isometry, the compatibility conditions on the folding amplitudes between neighboring vertices impose linear, homogeneous constraints among local isometries $\mathcal{V}^a(\mathbf{n})$ across different cells. In the Supplementary Materials we show explicitly how the constraint on each vertex in the unit cell can be represented as a corresponding row in a $4 \times 4$ matrix, which we denote the folding \emph{compatibility matrix}. The constraint equations are then written as:


\begin{align}\mathbf{C(\mathbf{z})}\ket{\mathcal{V}}=\mathbf{0}
\label{eq:compatibility}
\end{align}

\noindent where $\ket{\mathcal{V}}=(\mathcal{V}^a,\mathcal{V}^b,\mathcal{V}^c,\mathcal{V}^d)^T$ is the vector representing folding amplitudes of four vertices in a unit cell, and 

\begin{widetext}
\begin{align}
\mathbf{C}(\mathbf{z})=\begin{pmatrix}
X_1+X_2-X_3-X_4 & -X_1+z_1^{-1}X_3 & 0 & -X_2+z_2^{-1}X_4 \\
-X_1+z_1X_3 & X_1-X_2-X_3+X_4 & X_2-z_2^{-1}X_4 & 0 \\
0 & X_2-z_2X_4 & -X_1-X_2+X_3+X_4 & X_1-z_1X_3 \\
-X_2+z_2X_4 & 0 & X_1-z_1^{-1}X_3 & -X_1+X_2+X_3-X_4
\end{pmatrix} .
\label{eq:Cmat}
\end{align}
\end{widetext}

\noindent The $X_i$'s in\myeq{Cmat} are folding coefficients given by
\begin{align}
    X_i=\hat{\mathbf{r}}_{i+1}\cdot(\hat{\mathbf{r}}_{i+2}\times\hat{\mathbf{r}}_{i+3})/r_{i}
    \label{eq:Xs}
\end{align}
 where $\mathbf{r}_{i}$'s are edge vectors emanting out counterclockwise from vertex $a$, defined as in\myfig{figure1} (A). The subscript labels them in a counterclockwise order and is defined modulo $4$ since there are $4$ edges around each vertex. We emphasize that\myeq{Cmat} is obtained in a uniform reference state, while in a generic case where vertices with the same label across different unit cells are not in the same state, those coefficients are no longer defined globally and should depend explicitly on the cell index as well.

\subsection{Symmetries and topological class}

The $\mathbf{C}(\mathbf{z})$ defined in\myeq{Cmat} is generalized Hermitian in the sense that
\begin{align}
\mathbf{C}(\mathbf{z})^T=\mathbf{C(1/\mathbf{z})},
\end{align}

\noindent with $T$ denoting the matrix transpose. This reduces to the well-known Hermitian property $\mathbf{C}(\mathbf{z})^T=\mathbf{C}(\bar{\mathbf{z}})=\overline{\mathbf{C(\mathbf{z})}}$ inside the Brillouin zone, where $1/\mathbf{z}=\Bar{\mathbf{z}}$ with the overbar denoting complex conjugation. Hence, this matrix that describes the allowed linear isometries of an origami sheet can be treated as a Hamiltonian describing the quantum dynamics of a
periodic (crystalline) system. 
For our purpose, due to a spatial inversion symmetry in the parallelogram-based origami system, we adopt the topological classification for centrosymmetric systems introduced in~\cite{PhysRevB.96.155105}. 

We now consider explicitly the symmetries of this Hamiltonian. If we permute the opposite vertices in each parallelogram face, we exchange the mountains and valleys of the crease pattern, producing a global minus sign to $\mathbf{C}(\mathbf{z})$. Also, since the directions of the two lattice vectors are flipped, the generalized wavevector $\mathbf{z}$ becomes $1/\mathbf{z}$. In the end, we obtain a spatial inversion symmetry in the parallelogram-based origami lattice
{\begin{align}
    \mathbf{P}\mathbf{C}(\mathbf{z})\mathbf{P}^{-1}=-\mathbf{C}(1/\mathbf{z})
    \label{eq:inversion_symmetry}
\end{align}}

\noindent
where the permutation/inversion operator $\mathbf{P}$ has the form
\begin{align}
    \mathbf{P}=\begin{pmatrix}
0&0&1&0\\
0&0&0&1\\
1&0&0&0\\
0&1&0&0
\end{pmatrix}.
\end{align}
Obviously, we have $\mathbf{P}^2=\mathbb{I}$. Inspired by classification of quantum operators, we now introduce  the complex conjugation operator $\mathbf{K}$ (which, by definition, can be applied to any complex vector to produce its complex conjugate), we obtain an anti-unitary operator~\cite{peskin2018introduction} $\mathbf{KP}$ that obeys $(\mathbf{KP})^2= \mathbf{K}^2\mathbf{P}^2=\mathbb{I}$ since $\mathbf{P}$ is purely real and therefore commutes with $\mathbf{K}$, which also squares to the identity. Inside the Brillouin zone, we have
\begin{multline}
    (\mathbf{KP})\mathbf{C}(\mathbf{z})(\mathbf{KP})^{-1}=\mathbf{K}(\mathbf{P}\mathbf{C}(\mathbf{z})\mathbf{P})\mathbf{K}\\
    =-\mathbf{K}\mathbf{C}(\bar{\mathbf{z}})\mathbf{K}=-\mathbf{C}(\mathbf{z}),
    \label{eq:particle_hole}
\end{multline}

\noindent which corresponds to the particle-hole symmetry introduced in the AZ+$\mathcal{I}$ classification of centrosymmetric systems~\cite{PhysRevB.96.155105}.

Next, we show that there is also a chiral symmetry. Take $\theta(z_1)=\arg{(X_1-X_3z_1})$ and construct a unitary matrix as below which acts as a mirror symmetry in the reciprocal space
\ {\begin{align}
    \mathbf{M}_1(z_1)\equiv\begin{pmatrix}
e^{i\theta}& & & \\
 &e^{-i\theta}& & \\
 & &e^{-i\theta}& \\
 & & & e^{i\theta}
\end{pmatrix} 
\end{align}}
\noindent and we have 
\begin{align}
    \mathbf{M}_1(z_1)\mathbf{C}(z_1,z_2)\mathbf{M}_1(z_1)^{-1}=\mathbf{C}(\overline{z_1},z_2).
\end{align}

Similarly, take $\phi(z_2)=\arg(X_2-X_4z_2)$ and we have another unitary matrix
\begin{align}
    \mathbf{M}_2(z_2)\equiv\begin{pmatrix}
e^{i\phi}& & & \\
 &e^{i\phi}& & \\
 & &e^{-i\phi}& \\
 & & & e^{-i\phi}
\end{pmatrix} 
\end{align}
\noindent which satisfies
\begin{align}
    \mathbf{M}_2(z_2)\mathbf{C}(z_1,z_2)\mathbf{M}_2(z_2)^{-1}=\mathbf{C}(z_1,\overline{z_2}).
\end{align}

The composition of three unitary operators $\mathbf{P}\mathbf{M}_1\mathbf{M}_2$ is still unitary (where we suppress the arguments $z_{1,2}$ in $\mathbf{M}_{1,2}$ for now), and direct calculation shows that it squares to the identity
\begin{align}
    (\mathbf{P}\mathbf{M}_1\mathbf{M}_2)^2=\mathbb{I}
\end{align}

\noindent and its action on the $\mathbf{C}(\mathbf{z})$ is given by
\begin{multline}
    (\mathbf{P}\mathbf{M}_1\mathbf{M}_2)\mathbf{C}(\mathbf{z})(\mathbf{P}\mathbf{M}_1\mathbf{M}_2)^{-1}\\
    =\mathbf{P}\mathbf{C}(\overline{\mathbf{z}})\mathbf{P}^{-1}=-\mathbf{C}(\mathbf{z})
    \label{eq:chiral}
\end{multline}

\noindent which is exactly the chiral symmetry. Composing the chiral symmetry here and particle-hole symmetry\myeq{particle_hole} we have above, we obtain an anti-unitary operator \begin{align}
    \mathbf{T}\equiv \mathbf{PM}_1\mathbf{M}_2\mathbf{KP}
\end{align} with $\mathbf{T}^2=\mathbb{I}$ that serves as the PT symmetry and leaves $\mathbf{C}(\mathbf{z})$ invariant:
\begin{align}
\label{eq:time_reversal}
\mathbf{T}\mathbf{C}(\mathbf{z})\mathbf{T}^{-1}=\mathbf{C}(\mathbf{z}).
\end{align}

With all three symmetries\myeq{particle_hole},\myeq{chiral},\myeq{time_reversal} and the positive signatures, our system is categorized into the BDI class~\cite{PhysRevB.96.155105}, which has the sign of Pfaffian as its 0-D topological charge. For an anti-symmetric $2n\times 2n$ matrix $A=(a_{ij})$, the Pfaffian is defined to be
\begin{align}
    \mathrm{Pf}\ A=\frac{1}{2^nn!}\sum_{\sigma\in S_{2n}}\mathrm{sgn}(\sigma)\prod_{i=1}^na_{\sigma(2i-1),\sigma(2i)}
\end{align}
where $S_{2n}$ is the symmetric group of order $(2n)!$ and sgn$(\sigma)$ is the signature of the permutation $\sigma$.
Note that our $\mathbf{C}(\mathbf{z})$ is not anti-symmetric, even when $\mathbf{z}$ is inside the Brillouin zone. To resolve this apparent discrepancy, we can perform a change of basis and render $\mathbf{C}(\mathbf{z})$ into an anti-symmetric form. To reach this goal, consider
\begin{align}\mathbf{U}\equiv\mathbf{P}^{1/2}=\frac{1}{2}\rb{\mathbb{I}+\mathbf{P}}+\frac{i}{2}\rb{\mathbb{I}-\mathbf{P}}
\end{align}

\noindent which is unitary. We then do a unitary transformation $\Tilde{\mathbf{C}}(\mathbf{z})\equiv\mathbf{U}\mathbf{C}(\mathbf{z})\mathbf{U}^{-1}$, and it follows from direct calculation that\myeq{inversion_symmetry} becomes 
\begin{align}\Tilde{\mathbf{C}}(\mathbf{z})=-\Tilde{\mathbf{C}}^T(\mathbf{z})
\end{align}
\noindent which shows that $\Tilde{\mathbf{C}}(\mathbf{z})$ is now anti-symmetric and hence the definition of Pfaffian is justified. Denote by $\{\Tilde{c}_{ij}\}_{i,j=1}^4$ the entries of $\Tilde{\mathbf{C}}(\mathbf{z})$. These entries clearly depend on $\mathbf{z}$ but here we omit it for convenience. We have
\begin{multline}\mathrm{Pf}\ \Tilde{\mathbf{C}}(\mathbf{z})=\tilde{c}_{12}\tilde{c}_{34}-\tilde{c}_{13}\tilde{c}_{24}+\tilde{c}_{14}\tilde{c}_{23}\\
    =X_1X_3(z_1-1)^2/z_1-X_2X_4(z_2-1)^2/z_2
    \label{eq:Pfaffian}
\end{multline}
\noindent which is true for general complex wavevector $\mathbf{z}$ not necessarily inside the Brillouin zone. According to a property of Pfaffian that its square is equal to the determinant, we have
\begin{align}
    (\mathrm{Pf}\ \Tilde{\mathbf{C}}(\mathbf{z}))^2=\det{\Tilde{\mathbf{C}}(\mathbf{z})}=\det{\mathbf{C}(\mathbf{z})}.
    \label{eq:pfaffian_square}
\end{align}

Notice that due to the chiral symmetry\myeq{chiral}, the eigenvalues of $\mathbf{C}(\mathbf{z})$ are always paired $\pm\lambda$, thus $\det{\mathbf{C}(\mathbf{z})}$, which is equal to the product of all its eigenvalues, is non-negative. Hence, the Pfaffian on the left-hand side of\myeq{pfaffian_square} is purely real inside the Brillouin zone and its sign serves as a topological charge for classifying lattices.

\begin{widetext}

    \begin{figure*}

    \centering
    \includegraphics[width=\textwidth]{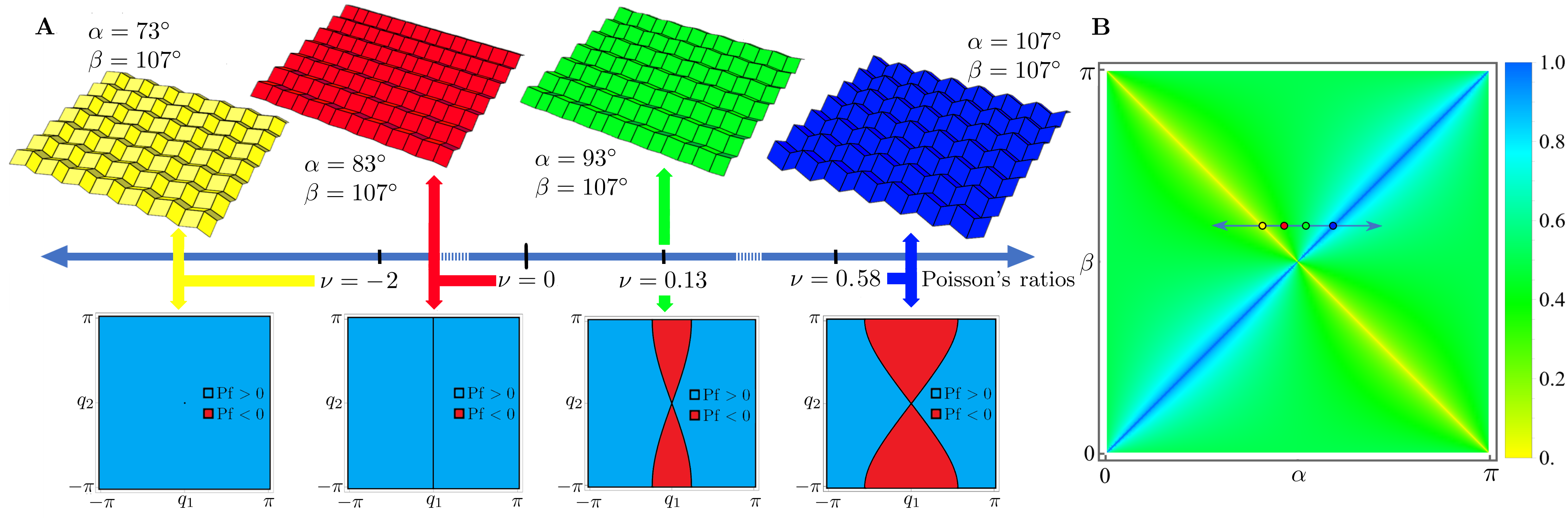}
    \caption{(A) Different Morph patterns (top) have rigid folding modes with different Poisson's ratios $\nu$.
    For auxetic systems ($\nu<0$) the topological invariant Pfaffian described in the main text has the same sign across the entire Brillouin zone (bottom), and the only bulk zero modes are at the origin, indicating long-wavelength behavior.
    For anauxetic systems ($\nu>0$) the invariant changes sign within the Brillouin zone,
    indicating doubly degenerate lines of zero modes (black), 
    such that low-energy response can occur at short wavelengths. The transition between the two cases occurs at $(\nu=0)$ with a vertical line of zero modes.
    (B) Each origami sheet has a nonlinear folding mode, as shown in Fig.\ref{fig:figure1}. This mode changes the Poisson's ratio and hence the presence of topological modes. The plot shows the fraction of configurations for a given sheet that have topological modes, ranging from 0 (Miura) to 1 (Eggbox). The blue arrow indicates our transition in (A) from the Miura (left) to the Eggbox (right), with four points corresponding to the four sheets with same colors in (A).
} 
    \label{fig:figure2}
\end{figure*}
\end{widetext}

\subsection{Topological classes determined by geometric factors and lattice Poisson's ratio}

Depending on whether the Pfaffian changes its sign in the Brillouin zone, parallelogram-based origami sheets can be categorized into two distinct classes:  conventional and  topological. The conventional sheets are those with a uniform sign of the Pfaffian inside the Brillouin zone, except for the origin where the determinant always vanishes. The topological sheets are those for which the Pfaffian changes signs  between two regions in the Brillouin zone separated by a topologically protected one-dimensional line as shown in\myfig{figure2}. Because the Pfaffian is the square root of the determinant, the determinant has a double root along this line, indicating doubly degenerate pairs of topologically protected zero modes.

Inside the Brillouin zone,\myeq{Pfaffian} becomes
\begin{align}
\label{eq:pfaffian_bz}\mathrm{Pf}\ \Tilde{\mathbf{C}}(\mathbf{q}) = 4\left(X_1X_3\sin^2\left(\frac{q_1}{2}\right)-X_2X_4\sin^2\left(\frac{q_2}{2}\right)\right).
\end{align}

As shown in~\cite{doi:10.1073/pnas.2202777119}, the lattice Poisson's ratio (the negative ratio of strains in the two lattice directions when the folding mode is activated) is  

\begin{align}
\label{eq:Poisson's_ratio}\nu=\frac{|l_2|^2}{|l_1|^2}\frac{X_2X_4}{X_1X_3}.
\end{align}

\noindent From these two expressions, it immediately follows that the existence of lines of zero modes in the Brillouin zone is completely determined by the sign of the Poisson's ratio. The Poisson's ratio of a Miura-ori shown in\myfig{figure1} is well-known to be always negative. Hence, the Pfaffian never flips its sign in the Brillouin zone and the only bulk zero mode occurs trivially at the origin with $q_1=q_2=0$. 

In contrast with the Miura, the Eggbox sheet\myfig{figure2}, always has a positive Poisson's ratio and thus always has doubly degenerate lines of zero modes. The Eggbox family thus fits into the category of topological lattices. For a generic Morph pattern, since its Poisson's ratio can take both positive and negative values \cite{PhysRevLett.122.155501}, its topological class depends not only on the unit cell geometry, but also on the particular spatial embedding. Using the Morph family, we are able to transit from a purely conventional lattice (a Miura) to a purely topological one (an Eggbox) as in\myfig{figure2}.

Four sample lattices, one from the Miura, two from the Morph, and one from the Eggbox family are shown in\myfig{figure2} to demonstrate a topological transition. As $\nu$ varies along the axis, lines of zero modes emerge vertically in the Brillouin zone at $\nu=0$, and, for a generic $\nu>0$, form an ``X" shape which gradually flattens and finally merges into a single horizontal line of zero modes $q_2=0$ at $\nu=+\infty$.  

In the next section we illustrate in detail how the topological transition fundamentally changes the mechanical response of the sheet under external loads. 

\section{Theoretical Analysis of Mechanical Response}
\label{sec:response}

\subsection{Description of deformation via vertex displacements and change of dihedral angles}

To investigate how these topological modes determine the response of an origami sheet to an external load, we adopt another formalism to describe deformation. In previous sections we use the $4\times4$ compatibility matrix\myeq{Cmat} to ensure geometric compatibility between neighboring vertices. It has the advantage of simplifying the description of isometric deformations to the point that the fundamental symmetries that govern the topological class can be discerned. However, such formulations cannot capture generic loading conditions in which panels are stretched as well as bent. To address this, we now describe a deformation of the $\mathbf{n}^\textrm{th}$ unit cell by a $12$-D vector $\mathbf{u}(\mathbf{n})$. This vector is obtained by concatenating four 3-D vectors, each representing the displacement of one of the cell's vertices~\cite{Rocklin2020}. Again, we assume a z-periodicity for the displacement vector:
\begin{align}
\mathbf{u}_{\mathbf{z}}(\mathbf{n}) \equiv \mathbf{u}_{\mathbf{z}}(n_1,n_2) \equiv z_1^{n_1}z_2^{n_2}\mathbf{u}_{\mathbf{z}}(\mathbf{0})\equiv\mathbf{z^n}\mathbf{u}_{\mathbf{z}}.
\label{eq:zperiodic}
\end{align}

In the case of an isometric deformation, it is possible to convert between vertex displacements and vertex folding, up to a uniform translation of the whole sheet, which do not induce any folding. Generic deformations to the origami sheet cost energy, which we can model by placing spring-like elements along the edges of origami panels. Because bending costs energy but does not stretch any edge, we also add a spring along a diagonal of each quadrilateral panel (the somewhat arbitrary choice of diagonal does not qualitatively impact the mechanical response, as verified by simulation).
To see the effect of infinitesimal displacements on those virtual springs, we introduce the rigidity matrix $\mathbf{R}(\mathbf{z})$~\cite{pratapa2018bloch}. In the literature, people also refer to this matrix as the compatibility matrix, but we prefer calling it the rigidity matrix in this paper, in order to avoid confusion with the $\mathbf{C}(\mathbf{z})$ matrix already defined in\myeq{Cmat}. The rigidity matrix $\mathbf{R}(\mathbf{z})$ relates, infinitesimally, the z-periodic displacements $\mathbf{u}_\mathbf{z}$ to resulting z-periodic extensions $\mathbf{e}_\mathbf{z}$ in bars (a compression being a negative extension) via 
\begin{align}
    \evec_\zvec = \rmat(\zvec) \uvec_\zvec,
\end{align}
and the dependence of $\mathbf{R}(\mathbf{z})$ on $\mathbf{z}$ is due to the existence of intercellular bars. Zero modes are by definition the null vectors of $\mathbf{R}(\mathbf{z})$ such that $\mathbf{R}(\mathbf{z})\mathbf{u}_\mathbf{z}=\mathbf{0}$, which also coincide with isometries. 

There are 12 bars in each unit cell, and so $\mathbf{e}_\mathbf{z}$ is also 12-D and $\mathbf{R}(\mathbf{z})$ is a $12\times 12$ square matrix. This means that our system is at the mechanical critical point where the number of degrees of freedom is equal to the number of constraints. For such systems, the Maxwell-Calladine index theorem~\cite{doi:10.1080/14786446408643668, CALLADINE1978161, Rocklin2020} ensures that a zero mode at $\zvec$ implies a state of self-stress (a force-balanced configuration in which tensions in bars joining at the same vertex result in a zero net force on the vertex) at $1/\zvec$. When $\zvec$ is complex, indicating exponential growth, the two modes are exponentially localized at opposite edges. In the special case of bulk modes ($\zvec$ real), one bulk zero mode at a certain wavevector corresponds to a state of self-stress at its opposite wavevector. For any triangulated origami sheets (not necessarily parallelogram-based), apart from this correspondence at opposite wavevectors, there is another hidden symmetry~\cite{doi:10.1073/pnas.2005089117} that establishes a duality between zero modes and states of self-stress at the same wavevector $\mathbf{z}$. Combining the hidden symmetry with the index theorem, we know that a zero mode at $\mathbf{z}$ give rises to another zero mode at $1/\mathbf{z}$. Therefore, zero modes appear in pairs on different edges, meaning that they are never polarized on one edge~\cite{PhysRevLett.116.135501, roychowdhury2018classification}, revealing that all two-dimensional sheets are topologically trivial within the Kane-Lubensky scheme of classifying topological mechanical modes~\cite{Kane2014}.

The determinant of $\mathbf{R}(\mathbf{z})$, when expanded into a Laurent polynomial in a proper gauge, has a palindromic form

\begin{align}
\label{eq:laurent}
    \det\mathbf{R}(\mathbf{z})=\sum_{m,n}c_{mn}z_1^{m}z_2^n
\end{align}

\noindent with $c_{mn}=c_{-m,-n}$. Here a gauge change means an overall division of $z_1^iz_2^j$ with some integers $i,j$ to make the coefficients match up. In particular, inside the Brillouin zone where $|z_1|=|z_2|=1$, zero modes appear in pairs at opposite real wavevectors. Furthermore, taking complex conjugates of both sides in\myeq{laurent} and using the palindromic property $c_{mn}=c_{-m,-n}$, we see that $\det\mathbf{R}(\mathbf{z})=\overline{\det\mathbf{R}(\mathbf{z})}$ which forces the determinant of $\mathbf{R}(\mathbf{z})$ to be real inside the 2-D Brillouin zone. 

Notice that even though\myeq{pfaffian_bz} is already shown to be real inside the Brillouin zone, the realness of\myeq{laurent} does not require the origami to be parallelogram-based at all. It means we can extend the topological classification of origami to any general triangulated periodic lattice by considering whether the sign of $\det{\mathbf{R}(\mathbf{z})}$ is conserved inside the Brillouin zone, and the conclusions we draw in this and following sections carry over to the case of generic triangulated lattices as well.

In order to take into account energy associated with folding along creases and bending of faces, which lifts the topological zero modes and turn them into soft modes,
we introduce another 12-D vector $\bm{\phi}$, representing the change of dihedral angles along creases. We treat the effect of bending a parallelogram face the same as folding along its diagonal crease, even though creasing folding and face bending are quantitatively different in practice. There are two main reasons to justify this simplification. One is that the topological signature of our system depends only on the rigidity matrix $\mathbf{R}(\mathbf{z})$ which reflects the behavior of the bars and does not distinguish crease folding from face bending. 
The second reason is that for real paperboard origami sheets, the energy cost of accommodating an external load via bending a face or folding along a crease are comparable, whereas stretching a face would require 
several orders of magnitude more energy~\cite{FILIPOV201726, schenk2011origami}.

The $\bm{\phi}$ induced by a z-periodic displacement is also z-periodic and we attach a $\mathbf{z}$ subscript to emphasize this dependence. By constructing an angular velocity field over the sheet as in~\cite{doi:10.1073/pnas.2005089117,doi:10.1073/pnas.2202777119}, we can find an explicit linear map $\mathbf{S}(\mathbf{z})$~\cite{pratapa2018bloch} that sends $\mathbf{u}_\mathbf{z}$ to $\bm{\phi}_\mathbf{z}$
\begin{align}
\bm{\phi}_\mathbf{z}=\mathbf{S}(\mathbf{z})\mathbf{u}_\mathbf{z}.\end{align}
\noindent This matrix also has the dimension $12\times12$ since each dihedral angle corresponds to a bar when there is no open boundary. The construction of this matrix is demonstrated in the Supplementary Materials.

\subsection{Quadratic energy functional and mechanical response}

In this section, we deduce a scaling law for the dominant response for topological sheets near soft modes under periodic boundary conditions\myeq{zperiodic}. From the dominant part, we can further predict that the effective stiffness of topological sheets should obey certain power laws, which are also verified numerically.

As the vertices displace, energy is required for the origami sheet to stretch and fold. In particular, under  periodic boundary conditions, we may assume the displacements to be modulated by a Bloch phase factor $e^{i\mathbf{q}}$, which is a special case of\myeq{zperiodic} with $z_1,z_2$ in the Brillouin zone. Correspondingly, the rigidity matrix $\mathbf{R}(e^{i\mathbf{q}})$ depends on this wavevector $\mathbf{q}$~\cite{evans2015lattice} and, for simplicity in notation, we denote it by $\mathbf{R}(\mathbf{q})$ from now on. The same change of notation applies to $\mathbf{S}$ as well. Under an external load, we introduce the quadratic energy functional $E_{\mathrm{tot}}$ to our system, induced by the periodic displacements $\mathbf{u}_\mathbf{q}$,
\begin{multline}
\label{eq:energy_function}
E_{\mathrm{tot}}=\sum_\textbf{q}  \,\,
\frac{1}{2}\textbf{u}^T_{-\textbf{q}}\textbf{S}^T(-\textbf{q})\textbf{K}_\textbf{f}\textbf{S}(\textbf{q})\textbf{u}_{\textbf{q}}\\+\frac{1}{2}\textbf{u}^T_{-\textbf{q}}\textbf{R}^T(-\textbf{q})\textbf{K}_\textbf{s}\textbf{R}(\textbf{q})\textbf{u}_{\textbf{q}}
\end{multline}

\noindent where $\textbf{K}_\textbf{s}$ is a diagonal matrix whose elements represent force constants of the bars and likewise $\textbf{K}_\textbf{f}$ is that of rotational springs associated with folding along creases. The exact values of elements in these two matrices in general depend on the particular geometry of the origami sheet as well as its specific embedding, but their orders of magnitude are uniform across a homogeneous sheet made up of the same material. Therefore, for convenience in discussion, we denote, respectively by $k_s$ and $k_f$, the representative orders of magnitude of diagonal elements in $\textbf{K}_\textbf{s}$ and $\textbf{K}_\textbf{f}$. For paperboard origami sheets, their numerical values are approximately $k_s=10^6\  (\mathrm{N\cdot m^{-1}})$ and $k_f= 0.1\ \mathrm{(N\cdot m)}$~\cite{FILIPOV201726, inproceedings}. For an origami sheet with average panel area $A \gtrapprox  10^{-4} \textrm{m}^2$, we may assume the unitless quantity $k_sA/k_f \gg 1$.

\begin{figure}
    \centering
    \includegraphics[width=8.5cm]{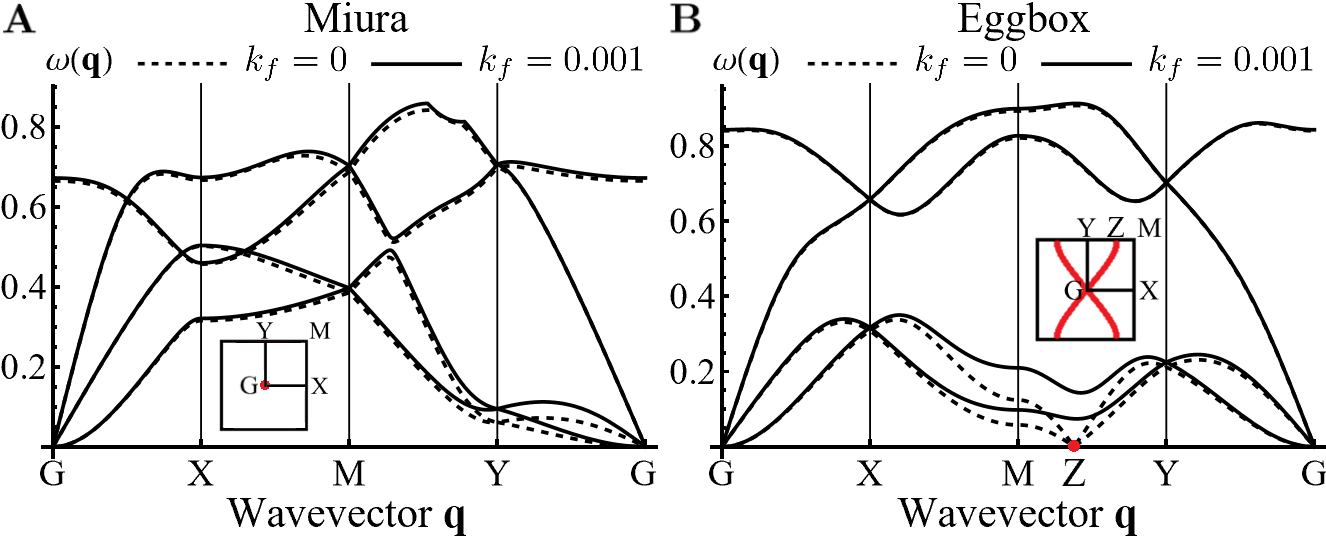}
    \caption{(A) Dispersion relations of a conventional sheet in the Miura family. The stretching stiffness $k_s$ is set to $1$ for simplicity. Dashed lines are for the case of no folding stiffness, $k_f=0$. Solid lines are for the case of small folding stiffness, $k_f=0.001$. There are three trivial zero modes (uniform translations in three directions) at the origin $G$ of the Brillouin zone. A non-zero folding stiffness does not significantly influence the dispersion relations. (B) Dispersion relations of a topological sheet in the Eggbox family. There are also three trivial zero modes at $G$, but two non-trivial topological zero modes appear between point $M$ and $Y$ (as they do for any path that intersects the red lines of isometries shown in the inset) in the absence of folding stiffness, as predicted by our analytical theory. In the absence of folding stiffness, the frequencies of the two acoustic modes scale linearly from the zero modes and verify the scaling law\myeq{frequency_scaling}. Increasing the folding stiffness lifts the frequencies of the two zero modes up by $O(k_f^{1/2})$.
} 
    \label{fig:figure_dispersion}
\end{figure}

The gradient of this energy function is related to the Fourier-transformed external force $\textbf{f}_\textbf{q}$ by 
\begin{multline}
\textbf{f}_\textbf{q}=\nabla_{\mathbf{u}_\textbf{q}}E_{\mathrm{tot}}=\textbf{S}^T(-\textbf{q})\textbf{K}_\textbf{f}\textbf{S}(\textbf{q})\textbf{u}_{\textbf{q}}\\+\textbf{R}^T(-\textbf{q})\textbf{K}_\textbf{s}\textbf{R}(\textbf{q})\textbf{u}_{\textbf{q}}\equiv\textbf{D}(\textbf{q})\textbf{u}_\textbf{q}.
\label{eq:force_balance}
\end{multline}
where we introduce the dynamical matrix $\textbf{D}(\textbf{q})$ relating the external load force to the resulting displacement via $\textbf{f}_\textbf{q}=\textbf{D}(\textbf{q})\textbf{u}_{\textbf{q}}$. Near the point $Z$ in\myfig{figure_dispersion}, with the mass of the vertices taken to be $1$, the frequencies of two acoustic modes follow
\begin{align}
    \omega^2(\mathbf{q})\approx O(1)(k_f/A)+O(1)k_s|\textbf{q}-\textbf{q}_{Z}|^2
    \label{eq:frequency_scaling}
\end{align}
where $\textbf{q}_Z$ is the point $Z$ in the Brillouin zone.

Solving\myeq{force_balance} to determine the response of an origami sheet to a load force involves inverting a $12\times12$ matrix $\mathbf{D}$ at each wavevector, which is unlikely to prove conceptually insightful. It is  now that the topological nature of the lattice provides qualitative insight into its mechanical response. In particular, one might assume the bending term in\myeq{force_balance} can be dropped as it is almost negligible  given $k_sA/k_f\gg 1$. However, this assumption breaks down if the sheet is topological, so that it has lines of soft modes where the stretching energy vanishes and the bending energy plays an important role. Therefore, the two terms are both small and comparable to each other in the case $\mathbf{q}$ stays very close to lines of soft modes. 

For finite-size sheets, the wavevectors allowed by the periodic boundary conditions consist of two-dimensional grid points in reciprocal space, none of which in general lie precisely on the lines of soft modes. However, for modes that are \emph{close} to this line of soft modes, there are two small quantities: the low bending modulus $k_f$ and the small distance in reciprocal space, which we may denote by $\delta\mathbf{q}$. The eigenvalues of the rigidity matrix thus should be on the order of $|\delta \mathbf{q}|$. It follows that, for the spectral decomposition of $\mathbf{R}^T(\mathbf{-q})\mathbf{R}(\mathbf{q})$ given by $(\lambda_{\mathbf{q},i},\mathbf{v}_{\mathbf{q},i})$, the lowest two eigenvalues $\lambda_{\mathbf{q},i}$ with $i=1,2$ should scale as $|\delta\mathbf{q}|^2$.

Denote by $\mathbf{u}_{\mathbf{q},i}\equiv\mathbf{v}_{\mathbf{q},i}^\dagger\mathbf{u}_\mathbf{q}$, $\mathbf{f}_{\mathbf{q},i}\equiv\mathbf{v}_{\mathbf{q},i}^\dagger\mathbf{f}(\mathbf{q})$ the projections of $\mathbf{u}_\mathbf{q}$ and $\mathbf{f}{(\mathbf{q})}$ into the direction of $\mathbf{v}_{\mathbf{q},i}$, we can use Cramer's rule~\cite{cramer1750introduction} to prove that the projection onto the first two eigenvectors dominates and so the following scaling law exists for $i=1,2$

\begin{align}
u_{\mathbf{q},i} \approx \frac{f_{\mathbf{q},i}}{O(1) (k_f/A)+O(1) k_s|\delta\mathbf{q}|^2}.
\label{eq:uscaling}
\end{align}

\noindent This means that modes close to these topological modes are excited to a great degree, owing to the low energetic cost.
The explicit derivation of the scaling law\myeq{uscaling} is shown in the Supplementary Materials.

The physical interpretation of this result is that for a topological sheet, the response $\mathbf{u}_\mathbf{q}$ is significantly higher near soft modes, and consists mostly of the two low-energy modes $\mathbf{u}_{\mathbf{q},1}$ and $\mathbf{u}_{\mathbf{q},2}$. 

It follows that the energy  $E(\mathbf{q})$ stored in a wavevector close to soft modes 
  $\mathbf{u}_\mathbf{q}$ is given by:
\begin{align}
\label{eq:energy_q}
    E(\mathbf{q})=\frac{1}{2}\mathbf{u}_\mathbf{q}^\dagger\mathbf{f}(\mathbf{q}) \approx \sum_{i=1,2}  \frac{f_{\mathbf{q},i}^2}{O(1) (k_f/A)+O(1) k_s|\delta\mathbf{q}|^2}.
\end{align}

Consequently, the total energy of the system has contributions mainly from modes near lines of soft modes. Intuitively it means that, since face folding costs significantly less energy than bar stretching, those modes near soft modes which do not change the lengths of any bars are energetically favored and activated to a greater degree in topological origami sheets. In the infinite-size limit, an analytical expression of the total energy can be deduced from\myeq{energy_q} with some proper approximations. In the Supplementary Materials, we show the total energy $E_{\mathrm{tot}}$ obeys the following power law:
\begin{align}
    \label{eq:power_law}
    E_{\mathrm{tot}}=
\int_{\mathrm{BZ}}E(\mathbf{q})d^2\mathbf{q}\sim\frac{|\mathbf{f}|^2}{\sqrt{k_s k_f/A}}.
\end{align}
The effective stiffness of the sheet, defined as $k_{\mathrm{eff}}=|\mathbf{f}|^2/2E_{\mathrm{tot}}$, consequently scales as $\sqrt{k_sk_f/A}$. With fixed $k_s,A$, for the conventional lattice, due to the absence of soft modes, its stiffness is independent of $k_f$ and scales as $k_s$ instead. Similar scaling laws of elastic moduli appear in other isostatic lattices~\cite{PhysRevE.83.011111, broedersz2011criticality}.

We also have a variant of the power law derived above for the special case of a vanishing Poisson's ratio, i.e., when the topological transition occurs. At $\nu=0$, the lines of soft modes merge into a single vertical line $q_1=0$ as shown in FIG.~\ref{fig:figure2}, and the lowest eigenvalue $\lambda_{\mathbf{q},1}$ is no longer proportional to $|\delta\mathbf{q}|^2$, but scales like $|\delta\mathbf{q}|^4$ instead. Correspondingly, equation\myeq{energy_q} becomes
\begin{align}
\label{eq:energy_q2}
    E(\mathbf{q})\approx \sum_{i=1,2}  \frac{f_{\mathbf{q},i}^2}{O(1) (k_f/A)+O(1) k_s|\delta\mathbf{q}|^4}
\end{align}
and the effective stiffness now scales like $k_s^{1/4}k_f^{3/4}$. Though we make an a strong assumption of working in the infinite-size limit, to derive these power laws analytically, they turn out to be obeyed by simulations of small, finite-size systems, as shown in\myfig{figure3}

At low folding stiffness $k_f/A\ll k_s$, due to the power law\myeq{power_law}, a topological sheet (as well as a transitional one) exhibits far less effective stiffness than a conventional one. Besides, since lines of soft modes emerge vertically at topological transition, the sheet at transition is more responsive to loads applied in the $\ell_2$ lattice direction. Generalizing this observation implies that, by calculating the direction of lines of soft modes which depends solely on the lattice Poisson's ratio $\nu$, we can qualitatively predict an origami sheet's mechanical response with respect to different external loads without explicit experimental or numeric simulations. 

Another consequence of having dominant response at finite-wavelength is that the response pattern is extremely jagged for topological lattices. Analogous to the diatomic vibration case where optical phonons are associated anti-phase oscillation for neighboring particles, the vertices in neighbored cells move in different directions in topological lattices, thus creating an extremely jagged response pattern, as opposed to smooth response in conventional lattices. 
We leave numerical verification of our claims in the next section.

\section{NUMERICAL RESULTS}
\label{sec:numerics}
\subsection{Periodic boundary condition}

\begin{widetext}

\begin{figure}[H]
    \centering
    \includegraphics[width=\textwidth]{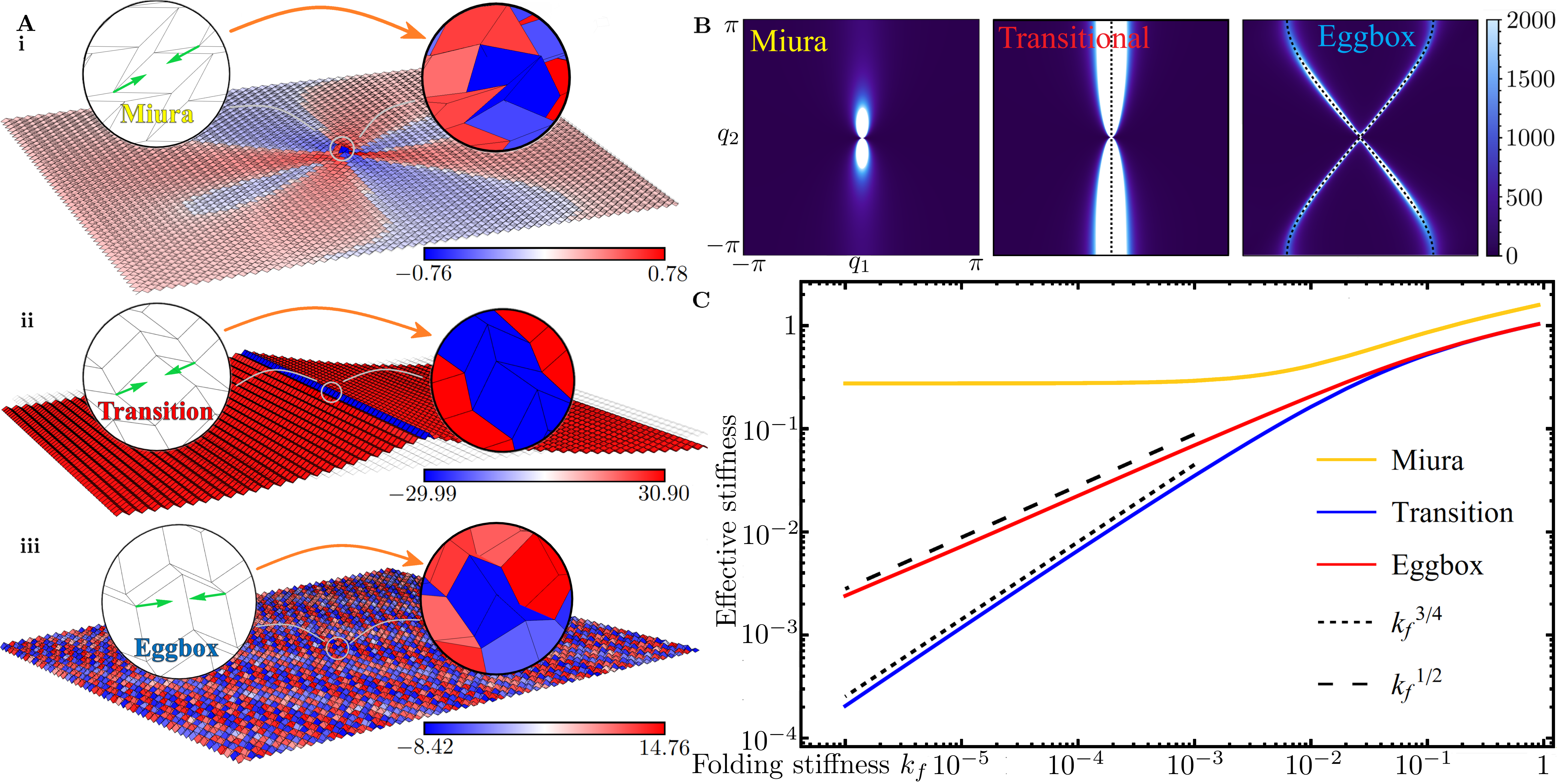}
    \caption{(A) Loading forces (green arrows) are applied in a linear, periodic model at the center of origami Morph sheets. Original positions of edges are shown in gray, with exaggerated linearly displaced vertices. The faces are colored according to the dilational component of local strain (red when expanding and blue when compressing). The topological sheets (the Eggbox and transitional one) exhibit larger deformation (in terms of root mean square of the strain) comparing to the conventional Miura sheet under the same dipole loading forces. (i) Conventional Miura-type sheets undergo a smooth, localized distortion. (ii) Sheets at the transition (zero Poisson's ratio) undergo compression that extends to the boundaries without decay. (iii) Topological Eggbox-type sheets undergo a sharply varying set of strains that largely cancel out, leading to little net displacement.
    (B) Magnitude of displacement is shown in reciprocal space in the Brillouin zone for the three lattices (from left to right, respectively, a Miura, a Morph at the transition point, and an Eggbox sheet). For visualisation convenience, all three plots are truncated so that high values in the white region are not shown explicitly, but near (dashed) lines of soft modes, the order of magnitude in the last two plots are significantly greater than that in the Miura plot (by a factor of order $1/k_f$). (C) Simulation of $50\times50$ finite-size systems with periodic boundary conditions. Effective stiffness of the Eggbox and the transitional lattice scales respectively as $k_f^{1/2}$ and $k_f^{3/4}$ with fixed $k_s=1$, verifying the power laws we derive in the main text. Also, notice at that high folding stiffness, the system no longer prefers the folding motion, and the effective stiffness is of the same order of $k_s=1$, reflecting a prevalence of bar stretching.}
    \label{fig:figure3}
\end{figure}
\end{widetext}

We begin by performing a linear calculation with periodic boundary conditions, the most analytically and numerically tractable scenario.
We numerically solve the reciprocal-space matrix equation\myeq{force_balance} for $50\times50$ systems such that cell $1$ and $51$ are identified in each lattice direction. As shown in\myfig{figure3} (A), we apply a force dipole to three systems: a Miura-ori , an Eggbox sheet and a Morph pattern with zero Poisson's ratio, thus at the transition point. 

In each case, the mechanical response is dominated by isometric modes that do not stretch any origami panels, marked by dashed lines in\myfig{figure3} (B). Because the Miura is topologically trivial, its only low-energy modes lie in the long-wavelength limit, and it behaves as a continuum solid sheet, with a smooth pattern of strains shown in\myfig{figure3} (A, i). In contrast, the presence of soft modes stretching across the Brillouin zone leads to a completely different response for the eggbox, with jagged strains such that neighboring unit cells might undergo opposite deformations in\myfig{figure3} (A, iii). For the transitional crease pattern in\myfig{figure3} (A, ii), the straight line of soft modes in reciprocal space leads to a simple, one-dimensional line of response in real space.

As shown in\myfig{figure3} (C), the differing patterns of spatial response also have profound effects on the effective stiffness of the sheet, the amount of force that is required to achieve a given amount of displacement. In the Miura's case, there is always long-wavelength stretching of panels, and consequently there is a finite sheet stiffness proportional to the panel stretching stiffness $k_s$ even in the absence of folding stiffness $k_f$. In contrast, as predicted by\myeq{power_law}, the Eggbox's lines of soft modes mean that the response will mix folding and stretching, so that the sheet's stiffness scales as  $k_s^{1/2} (k_f/A)^{1/2}$.  The transitional crease pattern in fact depends even more strongly on folding stiffness, with sheet stiffness scaling as $k_s^{1/4}(k_f/A)^{3/4}$ as predicted by\myeq{energy_q2}.

We point out that the plot\myfig{figure3} (A, iii) is expected to show an ``X" shape corresponding to the ``X'' shape of lines of soft modes in reciprocal space. The two ``X'' shapes should be perpendicular to each other at the origin. It is due to the fact that, in the reciprocal space, modes with wave vectors in the directions of four branches of the ``X'' shape are highly excited, and they give rise to nodal lines in the perpendicular directions, which form another ``X'' shape in real space. This observation is captured by results in the open boundary condition, as shown in\myfig{figure4}.

\subsection{Open boundary condition}
We use MERLIN~\cite{liu2018highly} to simulate origami deformation under open boundary conditions. The simulation settings and sheet parameters are specified in the Supplementary Materials.  MERLIN is a MATLAB package designed for performing nonlinear analysis of origami and its results agree well with real origami-based materials~\cite{MISSERONI2022101685}. 
\begin{widetext}

\begin{figure}[H]
    \centering
    \includegraphics[width=\textwidth]{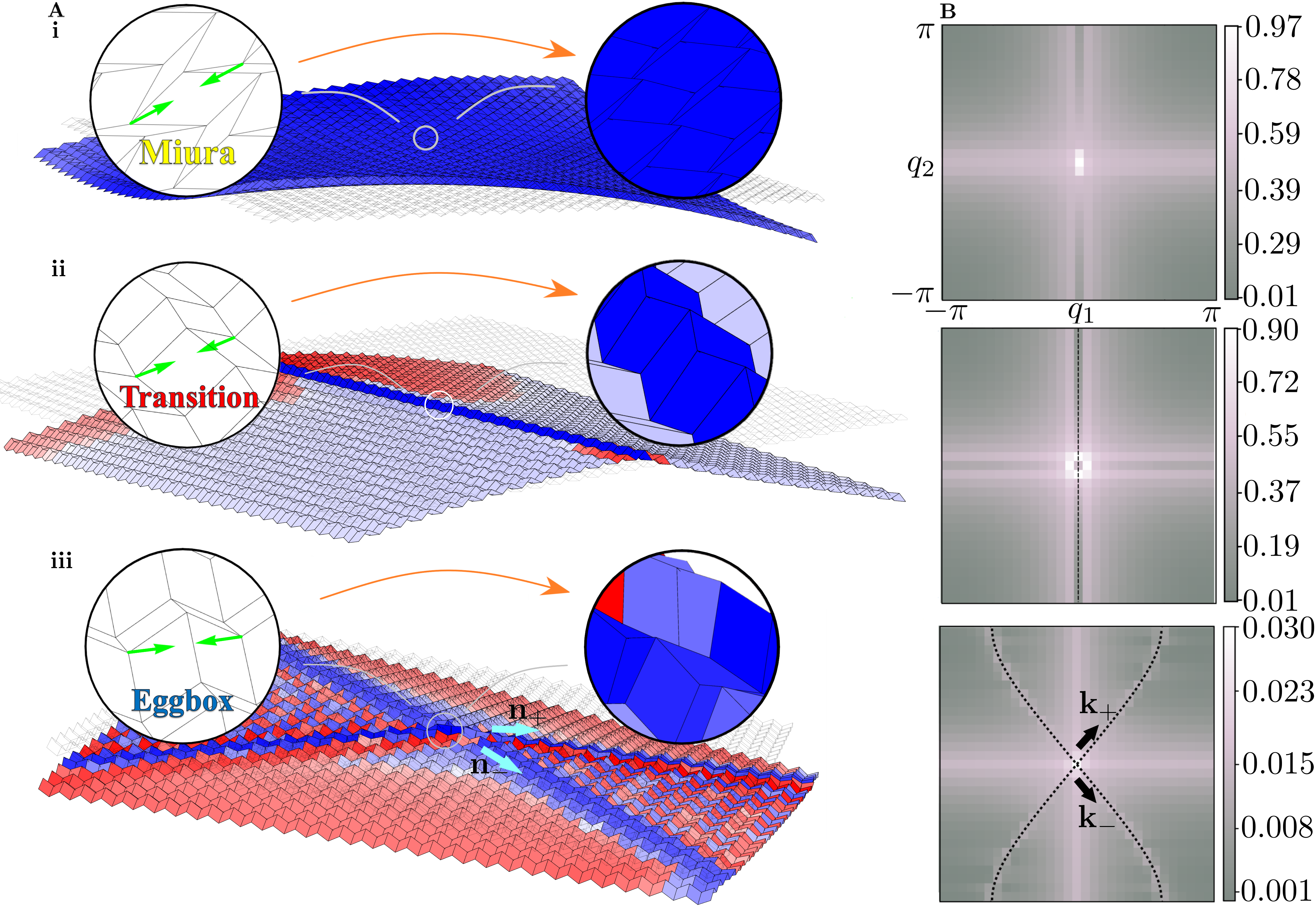}
    \caption{Counterpart to\myfig{figure3} in the case of open boundary conditions. (A) Spatial embeddings of the three sheets. The external load (green arrows) is infinitesimal, and resulted displacements are exaggerated. (i) a conventional Miura, (ii) a Morph at transition $\nu=0$, and (iii) a topological Eggbox, under a load at the center indicated by the pair of green arrows, with open boundary conditions. The panels are colored according to the dilation component of the in-plane strain tensor across the sheet. It is worth noting that the ``X''-shaped blue stripe in the real space response in (iii) corresponds to the ``X'' shape lines of soft modes in the reciprocal space by an orthogonality condition $\mathbf{k}_{\pm}\cdot\mathbf{n}_{\mp}=0$. (B) Excited modes in the reciprocal space line up with the theoretically predicted lines of soft modes (in black). The horizontal and vertical lines across the origin $q_{1,2}=0$ are also activated with the introduction of edge modes. 
    }

    \label{fig:figure4}
\end{figure}
\end{widetext}
With open boundary conditions, the introduction of edge modes to our system invalidates the power law established in last section. Nevertheless, bulk soft modes are well-reflected in the reciprocal space as shown in\myfig{figure4}. However,  modes at $q_1=0$ and $q_2=0$ are also possible under open  boundaries, with edge modes exponentially decaying as one moves into the bulk.
This appears in the reciprocal-space response of\myfig{figure4} (B) as a  ``$+$" shape at the origin. Excitation at finite wavelengths still makes the response pattern drastically different for conventional and topological lattices. The Miura, a paradigm of conventional lattices, exhibits a  smooth response, while the Eggbox in our family of topological lattices exhibits sharply jagged response. Finally, we mention that, even in an open-boundary system, the Eggbox family tends to be less stiff than the Miura one for a generic load, while the exact stiffness depends heavily on the particular combination of lattice Poisson's ratio and the direction of applied load. For instance, at $\nu=0$ where lines of soft modes appear vertically, the low-energy modes favored by the system are invariant in the $\ell_1$ direction, i.e., $q_1=0$. As a consequence, it is significantly easier to pull the lattice in the $\ell_2$ direction than in the $\ell_1$ direction. Indeed, this is implied by the vanishing Poisson's ratio, which indicates that deformation in one lattice direction does not excite deformation in the other lattice direction.

\section{Discussion}
\label{sec:discussion}

In this work, we show that different parallelogram-based origami sheets can have sharply distinct mechanical response, with smooth continuum fields in sheets and configurations with negative Poisson's ratios and jagged and spatially irregular deformations in those with positive Poisson's ratios. We show that this distinctive behavior is governed by a topological invariant, the Pfaffian, due to underlying symmetries of the system directly analogous to those present in quantum dynamics. 
We show that sheets for which this invariant takes on different values within the Brillouin zone
 which we refer to as \emph{topological lattices},
 possess topologically protected doubly degenerate lines of zero modes.
This theory provides a concrete way of classifying parallelogram-based origami sheets by direct calculation of this topological invariant.
Finally, using MERLIN simulation, we extend our results to systems with open boundary conditions.

Our results provide a topological perspective of controlling origami-inspired mechanical structures, with clearly testable experimental signatures. By choosing different crease geometries or by dynamically reconfiguring a single sheet, the lines of zero modes can be added and removed, with potentially dramatic implications for the origami's behavior. 
Recent advances have permitted the uniform excitation of origami folding modes~\cite{MISSERONI2022101685}, with the potential for modified methods to achieve and characterize the topological modes identified here. The doubly degenerate and protected nature of the modes suggests a potential realization of a holonomic (analog) computing regime identified in purely two-dimensional mechanical systems~\cite{fruchart2020dualities}.
Finally, the response of the origami systems may shed light on the behavior of their quantum-mechanical analog systems, as in~\cite{roychowdhury2018topology}.

\section{ACKNOWLEDGEMENTS}
The authors acknowledge financial support from the Army Research Office through the MURI program (\#W911NF2210219), the Office of Naval Research through the MURI program (\#N00014-20-1-2479) and through the National Science Foundation CAREER program (\# 2338492).

\bibliographystyle{unsrt}
\bibliography{bibliography}

\end{document}


\title{Supplementary materials}

\maketitle

\tableofcontents

\newpage

\section{Folding amplitude formalism and compatibility condition}

\subsection{Vertex folding}

We wish to describe infinitesimal changes to the orientations of the edges that are permitted by rigid folding motions. As we will see, these are essential to understanding deformations of the sheet that bend the faces without stretching them.

Consider a single origami vertex, with unit vectors $\hat{\mathbf{r}}_i$ emanating from the vertex along edge $i$, where $i$ is an index local to the vertex, so that $i=1$ corresponds to the edge pointing in the first lattice direction and $i$ increases counterclockwise from there. In a rigid motion, the dihedral angles between faces can change, but the sector angles between adjacent edges are fixed. One may therefore define the shape of the folding motion via $\{\phi_i\}$, infinitesimal changes to the dihedral angles. Consequently, one may show~\cite{doi:10.1073/pnas.2005089117} that the orientation of one face about another (set as reference) is given by a so-called ``angular velocity'' $\omega_j = \sum_{i=1}^j \phi_i \hat{\mathbf{r}}_i$, where the sum $j$ is over all edges one need to cross to reach the face from the reference one, such that any vector along the face $\mathbf{v} \rightarrow \mathbf{v} + \omega_j \times \mathbf{v}$. Consequently, we obtain the Belcastro Hull~\cite{BELCASTRO2002273} orientation compatibility condition around a vertex that the total change in angular velocity around the vertex is zero:

\begin{align}
    \sum_{i}\phi_i\hat{\mathbf{r}}_i=0.
    \label{eq:becastro_hull}
\end{align}

The summation is over all the edges emanating out from that vertex.
Note that in contrast, in the main text (see, in particular, Fig.~1), the edge vectors are defined globally and thus some end at rather than originate at a given vertex.

We are particularly interested in origami four-vertices, as are all vertices in the four parallelogram lattices considered in the main text.

The Belcastro-Hull condition then involves four dihedral angles and three independent constraints (the Cartesian components of the above equation), and it therefore follows from linear algebra that the solution is a 1-D linear subspace spanned by a single vector.
Every solution in this 1-D subspace is proportional to that basis vector and we call this proportionality factor the vertex folding amplitude $\mathcal{V}$ at that vertex.  

To obtain this basis vector from \myeq{becastro_hull}, we take an inner product of it with the vector $\hat{\mathbf{r}}_i\times\hat{\mathbf{r}}_{i+1}$ to get
\begin{align}
  \phi_i\hat{\mathbf{r}}_i\cdot(\hat{\mathbf{r}}_{i+1}\times\hat{\mathbf{r}}_{i+2})+\phi_{i+3}\hat{\mathbf{r}}_{i+3}\cdot(\hat{\mathbf{r}}_{i+1}\times\hat{\mathbf{r}}_{i+2})=0.
    \label{eq:becastro_hull2}
\end{align}
We remind the reader that the subscripts are defined modulo $4$ such that $i=5$ is equivalent to $i=1$. The (unnormalized) basis vector then yields the solution
\begin{align}
   \phi_i &=\mathcal{V} \zeta_i,
   \\
   \zeta_i&\equiv(-1)^i
\hat{\mathbf{r}}_{i+1}\cdot(\hat{\mathbf{r}}_{i+2}\times\hat{\mathbf{r}}_{i+3}).
    \label{eq:becastro_hull3}
\end{align}

\subsection{Face bending}

Consider now a parallelogram face initially in the plane, such that it has four edge vectors $\{\mathbf{r}_i\}$ with $\mathbf{r}_{i+2}= - \mathbf{r}_i$, where again $i$ is a local index for the face and defined modulo $4$. We remind the reader again that $\{\mathbf{r}_i\}$ are defined locally for each panel, and differ from how they are defined in previous section and in the main text.

\begin{figure}[H]
    \centering\includegraphics[width=0.8\textwidth]{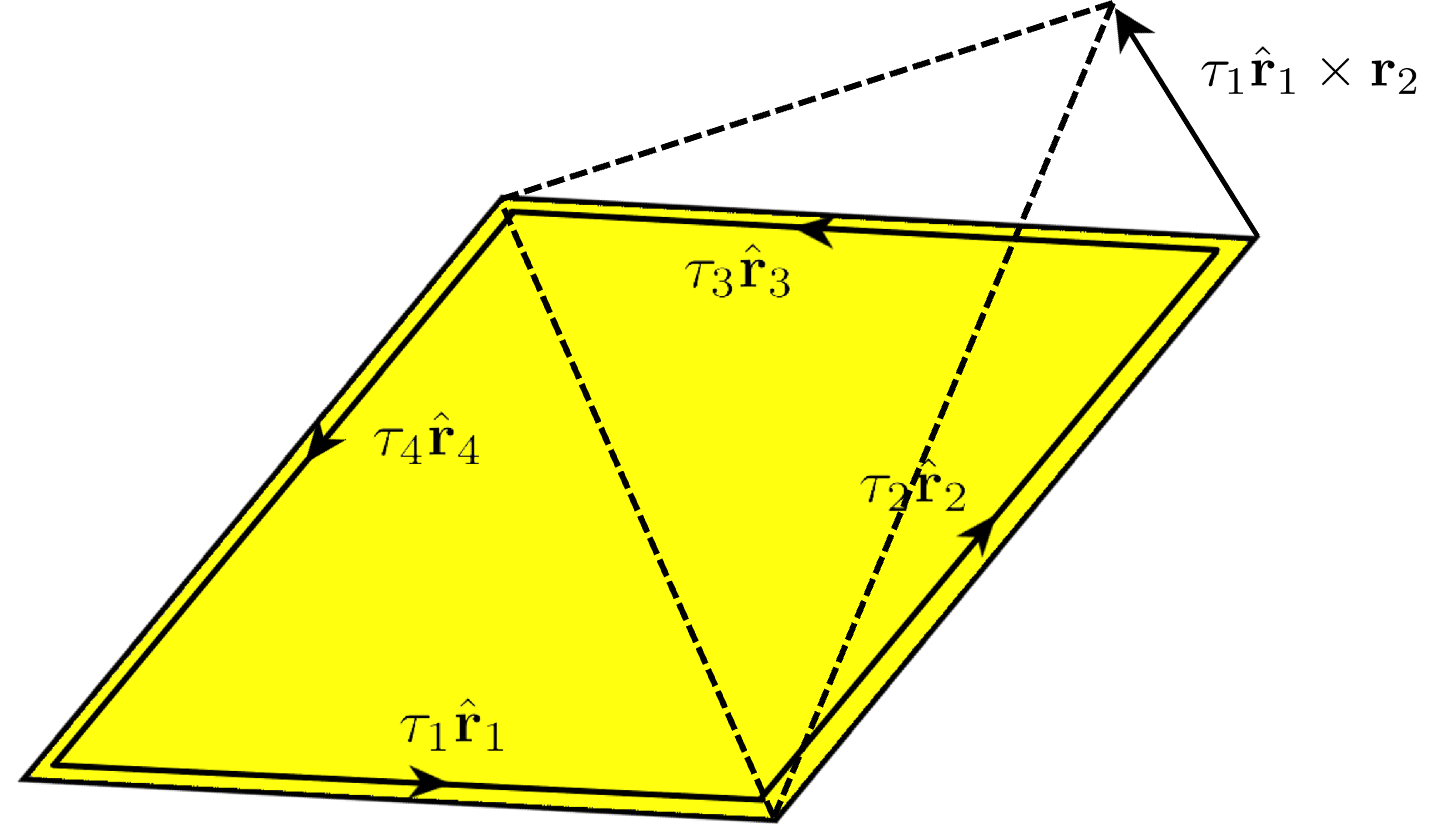}
    \caption{Two compatibility conditions need to be satisfied when a face is bent without being ripped. The orientation compatibility requires the local frame to preserve its orientation when travelling along the edges of the face in a full loop, while the position compatibility condition requires the origin of the local frame to return to its initial place.}
    \label{face_bending.png}
\end{figure}

We allow each face to bend along one of its diagonal by introducing along each edge a torsion that generates an angular velocity $\tau_i\hat{\mathbf{r}}_i$. Imagine a local frame that travels along the edges of a face and return to the starting location as indicated by the arrows in FIG. \ref{face_bending.png}. One compatibility condition is that the orientation of this frame should not change, and consequently we need
\begin{align}
    \sum_{i}\tau_i\hat{\mathbf{r}}_i=0.\label{eq:face_bending_orientation}
\end{align}

Apart from the orientation, the position of this frame (its origin) should return to the starting point as well, which requires
\begin{align}
    \tau_i\hat{\mathbf{r}}_i\times\mathbf{r}_{i+1}=\tau_{i-1}\hat{\mathbf{r}}_{i-1}\times\mathbf{r}_{i-2}.\label{eq:face_bending_position}
\end{align}

Combine the two compatibility conditions and the fact $\mathbf{r}_{i+2}= - \mathbf{r}_i$, we see that the solution of $\{\tau_i\}$ is again a 1-D linear space spanned by the vector $(-r_1,r_2,-r_3,r_4)^T$
\begin{align}
\tau_i=(-1)^i\mathcal{F}r_i
\end{align}
where the scalar $\mathcal{F}$ is called the face bending amplitude.

\subsection{Edge compatibility condition}

\begin{figure}[H]
    \centering\includegraphics[width=0.8\textwidth]{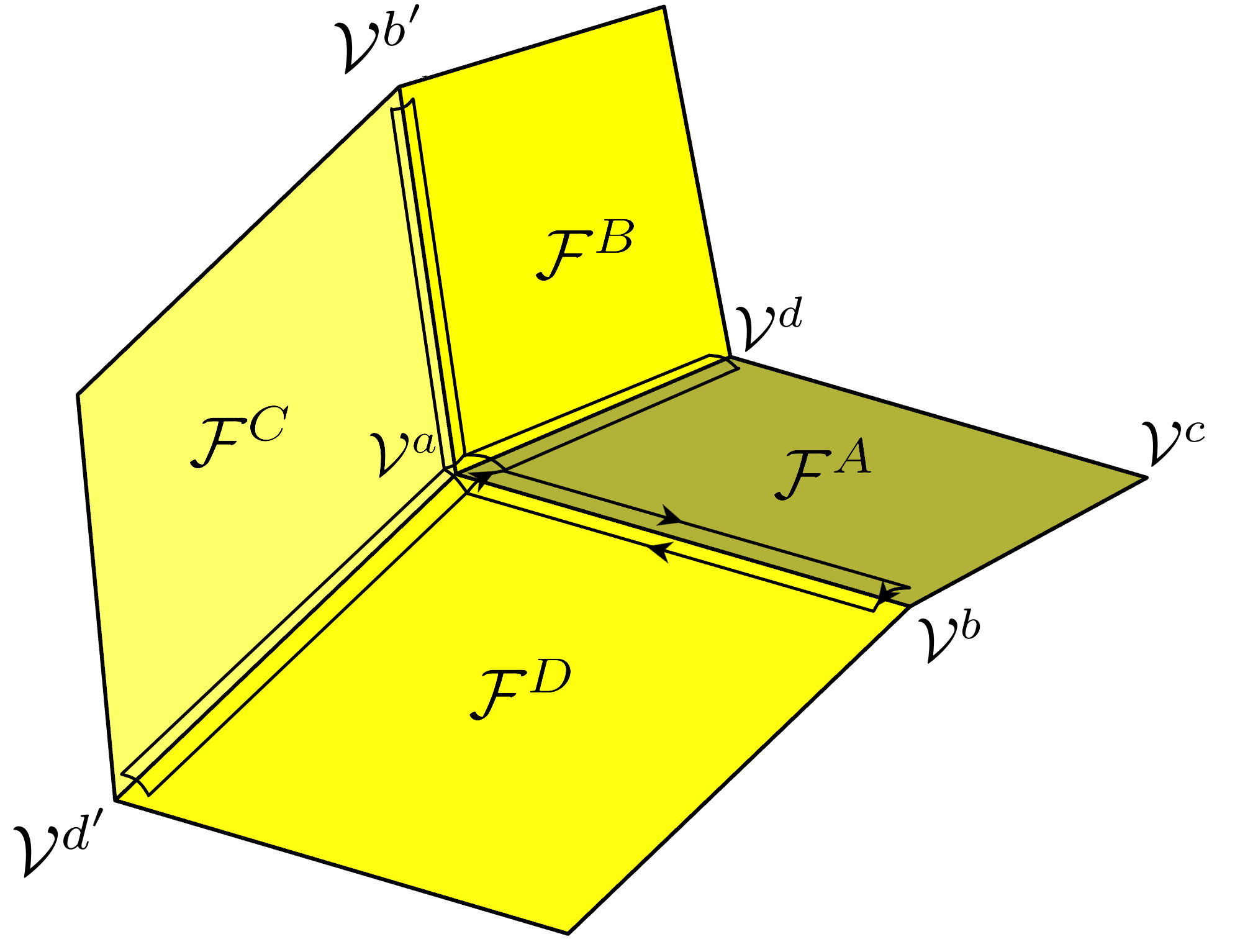}
    \caption{The orientation of a local frame should be preserved when it travels along the 16 path segments which form a ``cross'' around the vertex $a$. This compatibility condition consists of 4 edge compatibility conditions, among which we use black arrows to indicate one of them: 4 path segments constituting the edge compatibility condition along the edge between vertex $a$ and $b$. }
    \label{fig:compatibility_cross.png}
\end{figure}

In previous sections we define the vertex folding amplitude $\mathcal{V}$ and the face bending amplitude $\mathcal{F}$. From now on, we add superscripts to $\mathcal{V}^a$ and $\mathcal{F}^A$ to label the corresponding vertex $a$ and face $A$. We introduce another compatibility condition along every edge. This compatibility condition involves two of the vertex folding amplitudes and two of the face bending amplitudes. For instance, in\myfig{compatibility_cross.png}, if we travel along the rectangular loop  (indicated by four black arrows) around the edge connecting two vertices $a$ and $b$, we obtain a compatibility condition on this edge that preserves the orientation of the local frame. The position of this frame is always preserved since the two shorter sides of this rectangular loop (the two between face $A$ and $D$) are taken to be infinitesimally small, and no net displacement is generated. If  we call $\mathbf{r}_1^a$ the edge vector emanating from $a$ and reaching $b$, we have the following relation for the two vertex folding amplitudes $\mathcal{V}^a, \mathcal{V}^b$ and two face bending amplitudes $\mathcal{F}^A,\mathcal{F}^D$ 
\begin{align}
    \zeta^a_1\mathcal{V}^a-r_1^a\mathcal{F}^A-
    \zeta^b_3\mathcal{V}^b+r_3^b\mathcal{F}^D=0.
    \label{eq:edge_compatibility}
\end{align}
Notice that $\mathcal{F}^A$ and $\mathcal{F}^D$ have the same coefficient $r_1^a=r_3^b$ and opposite sign in\myeq{edge_compatibility}. Thus, divide by $r_1^a$ and rearrange to get
\begin{align}
    \frac{\zeta^a_1}{r_1^a}\mathcal{V}^a-
    \frac{\zeta^b_3}{r_1^a}\mathcal{V}^b=\mathcal{F}^A-\mathcal{F}^D.
    \label{eq:edge_compatibility2}
\end{align}
Due to symmetry, we have $\zeta_3^b=\zeta_1^a$, and so we are motivated to define the geometric factor $X_i^a=(-1)^i\zeta_i^a/r_i^a$ at vertex $a$, in terms of which\myeq{edge_compatibility2} is rewritten as
\begin{align}
    -X_1^a\mathcal{V}^a+X_1^a\mathcal{V}^b=\mathcal{F}^A-\mathcal{F}^D.
    \label{eq:edge_compatibility3}
\end{align}

We generalize this condition by writing
\begin{align}
    \frac{\zeta_i^a}{r_i^a}-\frac{\zeta_{i+2}^{a'}}{r_{i+2}^{a'}}=(-1)^iX_i^a\mathcal{V}^a-(-1)^{i+2}X_{i+2}^{a'}\mathcal{V}^{a'}=\mathcal{F}^A-\mathcal{F}^{A'}
    \label{eq:edge_compatibility4}
\end{align}
where any two adjacent faces $A$ and $A'$ join at an edge connecting the two vertices $a$ and $a'$.\myeq{edge_compatibility2} and\myeq{edge_compatibility3} are special cases of\myeq{edge_compatibility4} with specific labels $a'=b$ and $A'=D$. 

\subsection{``Cross'' compatibility condition}

The right-hand side of\myeq{edge_compatibility4} is the difference of two face bending amplitudes, and if we write down\myeq{edge_compatibility4} for all four edges around vertex $a$ and add them up as in\myfig{compatibility_cross.png}, the right-hand side cancels out. The left-hand side has linear terms of $
\mathcal{V}^{a},\mathcal{V}^{b},\mathcal{V}^{d},\mathcal{V}^{b'},\mathcal{V}^{d'}$ with coefficients being the geometric factors $X_i^{a,b,d,b',d'}$ defined at corresponding vertices. 
Because each face is a parallelogram, each vertex has edges pointing in the same four directions (and with the same four lengths), meaning that the geometric factors obtained by taking triple products of these directions are the same at every vertex up to order and sign. We can express each such factor in terms of those four factors at the $a$ vertex. For instance, using $\xi_3^b=\xi_1^a$ and $r_1^a=r_3^b$ in\myfig{compatibility_cross.png}, we know $X_1^a=X_3^b$, and that is exactly how we rewrite\myeq{edge_compatibility} into\myeq{edge_compatibility3}. Therefore, we turn all those geometric factors into the ones defined at the vertex $a$ and omit the superscript in $X_i$. 

This compatibility condition is known as the ``cross'' compatibility condition at vertex $a$, since the four edge conditions form a cross shape around vertex $a$. We sum up the four conditions, each given by\myeq{edge_compatibility4} with different labels
\begin{align}
    \sum_{(i,a')}\left(\frac{\xi_i^a}{r_i^a}-\frac{\xi_{i+2}^{a'}}{r_{i+2}^{a'}}\right)&=
\\ \nonumber
\sum_{(i,a')}\left((-1)^iX_i^a\mathcal{V}^a-(-1)^{i+2}X_{i+2}^{a'}\mathcal{V}^{a'}\right)&=\sum_{(A,A')}(\mathcal{F}^A-\mathcal{F}^{A'})=0.
\end{align}
Here, the summation $(i,a')$ is over the four directions $i$ and the four vertices $a'$ that are neighbors of vertex $a$ in the corresponding directions. The indices $(A,A')$ refer to the faces that are adjoining vertex $a$ and counter-clockwise and clockwise respectively from edge $i$. The final summation must be zero because all the face amplitudes cancel out. 
For example, for the vertex $a$ depicted in 
\myfig{compatibility_cross.png} the pair of labels $(i,a')$ take the values $\{(1,b),(2,d),(3,b'),(4,d')\}$ and correspondingly $(A,A')$ in the last summation takes $\{(A,D),(B,A),(C,B),(D,C)\}$. Writing those labels explicitly, we have
\begin{multline}
    (X_1\mathcal{V}^a-X_1\mathcal{V}^b)+(X_2\mathcal{V}^a-X_2\mathcal{V}^d)+(-X_3\mathcal{V}^a+X_3\mathcal{V}^{b'})+(-X_4\mathcal{V}^a+X_4\mathcal{V}^{d'})\\
    =(X_1+X_2-X_3-X_4)\mathcal{V}^a+(-X_1\mathcal{V}^b+X_3\mathcal{V}^{b'})+(-X_2\mathcal{V}^d+X_4\mathcal{V}^{d'})=0.
    \label{eq:first_row}
\end{multline}

We are particularly interested in the generalized Bloch modes, in which advancing in lattice direction $i$ multiplies the vertex (and face) amplitudes by some complex number $z_i$. In this way, we have, for example, $\mathcal{V}^{b} = z_1 \mathcal{V}^{b'}$ and $\mathcal{V}^{d}=z_2\mathcal{V}^{d'}$ in\myfig{compatibility_cross.png}. In the case that both $|z_i|=1$ these are periodic solutions that span the space of all periodic solutions. For generic $z_i$, they can represent solutions that are exponentially localized to different edges. For such a mode, by linearity if the compatibility equations are satisfied at every vertex in one unit cell, they are also satisfied throughout the sheet. As such, we can express the four constraints as a $\mathbf{z}$-dependent matrix equation via the vector $\ket{\mathcal{V}}$ consisting of amplitudes $\mathcal{V}^a,\mathcal{V}^b,\mathcal{V}^c,\mathcal{V}^d$:
\begin{align}\mathbf{C(\mathbf{z})}\ket{\mathcal{V}}=\mathbf{0},
\label{eq:cmat_ket}
\end{align}
where the compatibility matrix $\mathbf{C}(\mathbf{z})$ depends on the generalized wave number $\mathbf{z}$
\begin{align}
\mathbf{C}(\mathbf{z})=\begin{pmatrix}
X_1+X_2-X_3-X_4 & -X_1+z_1^{-1}X_3 & 0 & -X_2+z_2^{-1}X_4 \\
-X_1+z_1X_3 & X_1-X_2-X_3+X_4 & X_2-z_2^{-1}X_4 & 0 \\
0 & X_2-z_2X_4 & -X_1-X_2+X_3+X_4 & X_1-z_1X_3 \\
-X_2+z_2X_4 & 0 & X_1-z_1^{-1}X_3 & -X_1+X_2+X_3-X_4
\end{pmatrix} .
\label{eq:cmat}
\end{align}
\subsection{Symmetries and topological class}

As can be seen explicitly from the form above, this matrix $\mathbf{C}(\mathbf{z})$ is Hermitian inside the Brillouin zone where $1/\mathbf{z}=\Bar{\mathbf{z}}$, as
\begin{align}
    \mathbf{C}(\mathbf{z})^T=\mathbf{C}(1/\mathbf{z})=\mathbf{C}(\bar{\mathbf{z}})=\overline{\mathbf{C}(\mathbf{z})}.
\end{align}

Consider a spatial inversion symmetry given by
\begin{align}
    \mathbf{P}=\begin{pmatrix}
0&0&1&0\\
0&0&0&1\\
1&0&0&0\\
0&1&0&0
\end{pmatrix}.
\end{align}
\noindent  Again, from explicit calculation, we find that 
\begin{align}
    \mathbf{P}\mathbf{C}(\mathbf{z})\mathbf{P}^{-1}=-\mathbf{C}(1/\mathbf{z}).
\label{eq:inversion_symmetry}
\end{align}
If we compose it with the complex conjugation operator $\mathbf{K}$, we obtain an anti-unitary operator~\cite{peskin2018introduction} $\mathbf{KP}$ such that (within the Brillouin zone)
\begin{align}
    (\mathbf{KP})^2 &= \mathbf{K}^2\mathbf{P}^2 =\mathbb{I}\\
    (\mathbf{KP})\mathbf{C}(\mathbf{z})(\mathbf{KP})^{-1} &=\mathbf{K}(\mathbf{P}\mathbf{C}(\mathbf{z})\mathbf{P})\mathbf{K}=-\mathbf{K}\mathbf{C}(\bar{\mathbf{z}})\mathbf{K}=-\mathbf{C}(\mathbf{z})
    \label{eq:particle_hole}    
\end{align}
\noindent which corresponds to the particle-hole symmetry~\cite{Zirnbauer_1996, PhysRevB.55.1142} for a quantum mechanical system.

Take $\theta(z_1)=\arg{(X_1-X_3z_1})$, $\phi(z_2)=\arg(X_2-X_4z_2)$ and consider two mirror symmetries
\ {\begin{align}
    \mathbf{M}_1(z_1)\equiv\begin{pmatrix}
e^{i\theta}& & & \\
 &e^{-i\theta}& & \\
 & &e^{-i\theta}& \\
 & & & e^{i\theta}
\end{pmatrix},\ \     \mathbf{M}_2(z_2)\equiv\begin{pmatrix}
e^{i\phi}& & & \\
 &e^{i\phi}& & \\
 & &e^{-i\phi}& \\
 & & & e^{-i\phi}\end{pmatrix}
\end{align}}
\noindent which satisfy 
\begin{align}
    \mathbf{M}_1(z_1)\mathbf{C}(z_1,z_2)\mathbf{M}_1(z_1)^{-1}=\mathbf{C}(\overline{z_1},z_2)\\
    \mathbf{M}_2(z_2)\mathbf{C}(z_1,z_2)\mathbf{M}_2(z_2)^{-1}=\mathbf{C}(z_1,\overline{z_2})
\end{align}

Composing three operators $\mathbf{P}\mathbf{M}_1\mathbf{M}_2$ gives us another unitary operator which squares to identity
\begin{align}
    (\mathbf{P}\mathbf{M}_1\mathbf{M}_2)^2=\mathbb{I}
\end{align}
\noindent and satisfies the chiral symmetry
\begin{align}
    (\mathbf{P}\mathbf{M}_1\mathbf{M}_2)\mathbf{C}(\mathbf{z})(\mathbf{P}\mathbf{M}_1\mathbf{M}_2)^{-1}
    =\mathbf{P}\mathbf{C}(\overline{\mathbf{z}})\mathbf{P}^{-1}=-\mathbf{C}(\mathbf{z})
    \label{eq:chiral}
\end{align}

With the particle-hole and chiral symmetries available, we compose them to get another anti-unitary operator $\mathbf{T}$ with $\mathbf{T}^2=\mathbb{I}$ that leaves $\mathbf{C}(\mathbf{z})$ invariant
\begin{align}
    \mathbf{T}\mathbf{C}(\mathbf{z})\mathbf{T}^{-1}=\mathbf{C}(\mathbf{z})\label{eq:time_reversal}.
\end{align}

\noindent This corresponds to the time-reversal symmetry in quantum mechanics. In the end, since all three symmetries exist for our system, and the particle and time-reversal symmetries have positive signature (square to $\mathbb{I}$ instead of $-\mathbb{I}$), our system is categorized into the BDI class~\cite{Zirnbauer_1996,PhysRevB.55.1142}, which has the sign of Pfaffian as its 0-D topological charge~\cite{PhysRevB.96.155105}. The explicit construction of the Pfaffian is illustrated in the main text.

\section{Construction of the matrix $\mathbf{R}$ and $\mathbf{S}$}
\subsection{Construction of the matrix $\mathbf{R}$}

\begin{figure}[H]
    \centering\includegraphics[width=0.8\textwidth]{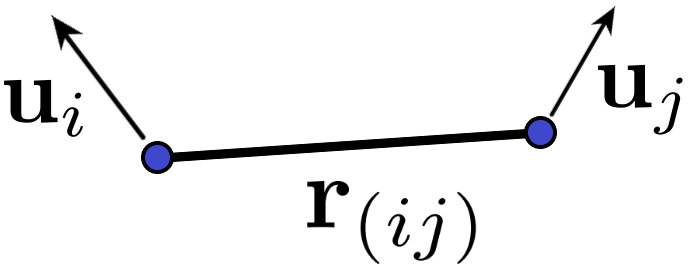}
    \caption{Two sites $i, j$, each with displacement $\mathbf{u}_i$ and $\mathbf{u}_j$, cause the length of the bar connecting the two sites to change. The infinitesimal extension/ compression of the bar is given by the projection of relative displacement of the two sites onto the direction of the bar.}
    \label{fig:bar_extension.png}
\end{figure}

The rigidity matrix $\mathbf{R}$ is easily obtained by projecting displacements onto the directions of the bars. For instance, given two sites $i$ and $j$ that are connected by a bar, and let $\mathbf{r}_{(ij)}$ be the vector joining the two sites from $i$ to $j$, as shown in\myfig{bar_extension.png}. If we displace $i,j$ by the vectors $\mathbf{u}_i,\mathbf{u}_j$, then infinitesimally, the change of length of  this bar is given by
\begin{align}
{e}_{(ij)}=(\mathbf{u}_j-\mathbf{u}_i)\cdot\hat{\mathbf{r}}_{(ij)}=\frac{(\mathbf{u}_j-\mathbf{u}_i)\cdot\mathbf{r}_{(ij)}}{r_{(ij)}}.
\end{align}

The ${e}_{(ij)}$ depends linearly on $\mathbf{u}_{(ij)}$. Therefore, by concatenating them respectively to form two vectors $\mathbf{e}$ and $\mathbf{u}$, there is a matrix $\mathbf{R}$ such that $\mathbf{R}\mathbf{u}=\mathbf{e}$. If each site is allowed to move freely in 3-D space, the dimensions of $\mathbf{R}$ are $N_b\times 3N$ where $N_b$ is the number of bars and $N$ is the number of sites in the system. 

If the system is periodic and the mode is $z$-periodic as described in the previous section, the matrix $\mathbf{R}$ depends on $\mathbf{z}$ explicitly so long as there are bars across different cells. The matrix $\mathbf{R}$ can be block-diagonlized with each block having dimension $n_b\times 3n$ where $n_b,n$ are now the numbers of bars and sites per unit cell. For our triangulated origami sheet, we have $n_b=12, n=4$, as shown in\myfig{unit_cell.png} below. 
\begin{figure}[H]
    \centering\includegraphics[width=0.8\textwidth]{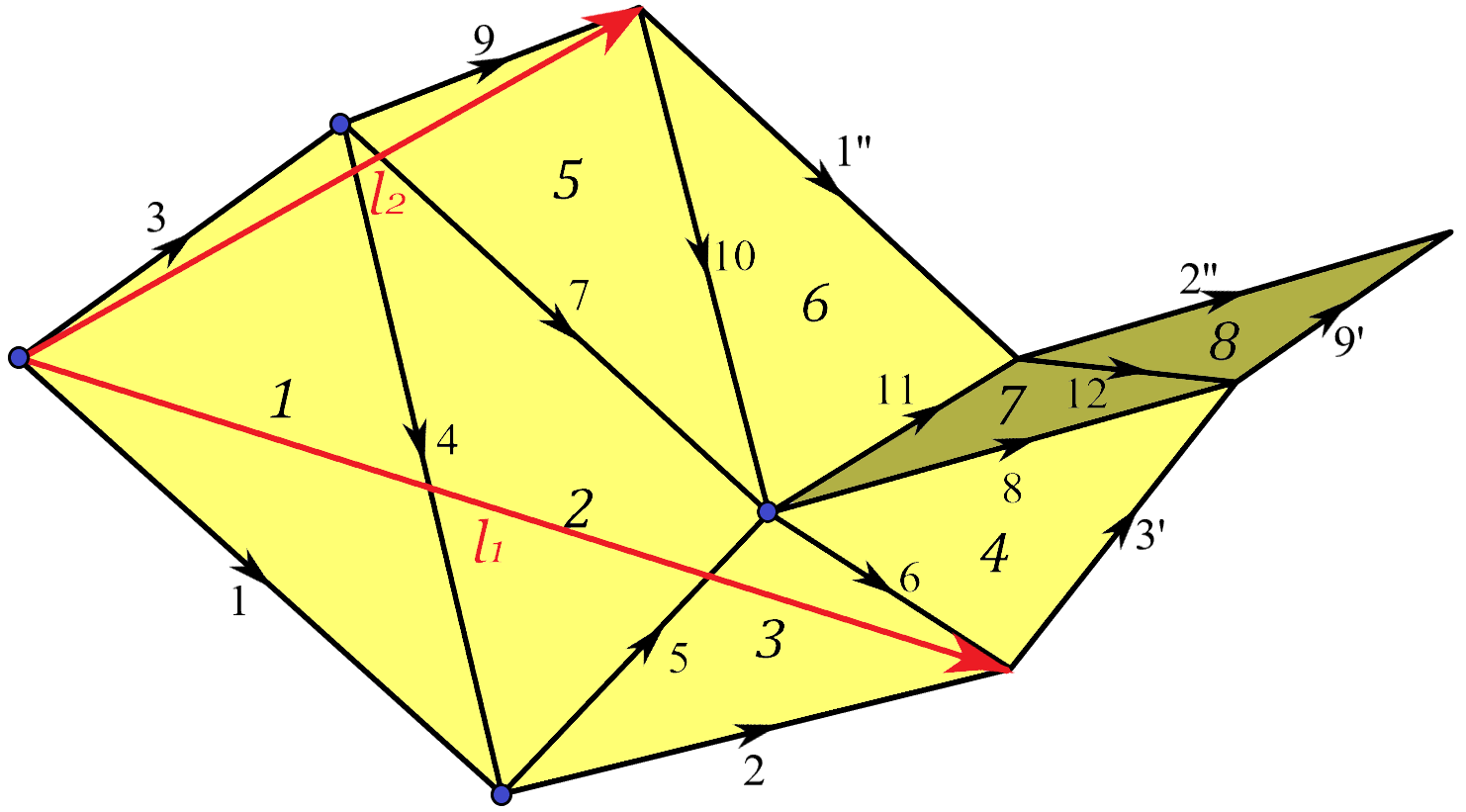}
    \caption{There are four vertices in every unit cell (indicated by circles), while the other nodes at the two ends of bars $1'',2'',3',9'$ belong to neighboring cells. The bars $1'',2'',3',9'$ also belong to neighboring cells, so there are exactly $12$ bars and $8$ triangulated faces per unit cell.} 
    \label{fig:unit_cell.png}
\end{figure}

\subsection{Construction of the matrix $\mathbf{S}$}

\begin{figure}[H]
    \centering\includegraphics[width=0.8\textwidth]{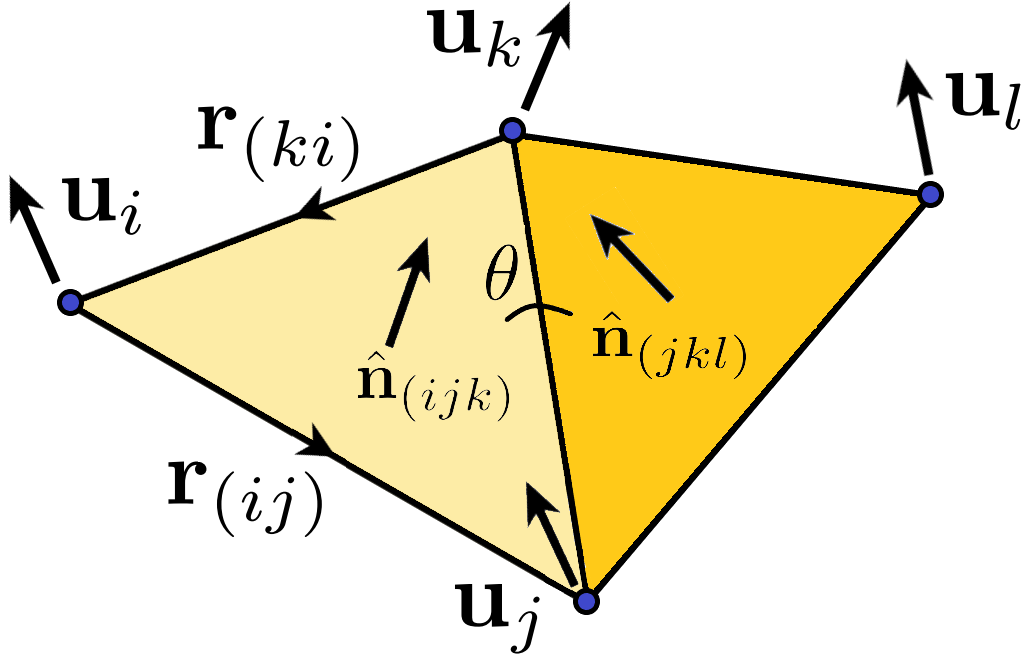}
    \caption{The displacements $\mathbf{u}_{i,j,k,l}$ at four sites indicated by circles induce two angular velocities $\bm{\omega}_{(ijk)}$ and $\bm{\omega}_{(jkl)}$. The difference of these two angular velocities causes folding along the bar connecting $j$ and $k$. } 
    \label{fig:Smat.png}
\end{figure}

Here we demonstrate how to construct the matrix $\mathbf{S}$ that maps the displacement vector $\mathbf{u}$ to the changes in dihedral angels $\bm{\phi}$. Using two adjacent panels as an example as in\myfig{Smat.png}, we first focus on the face bounded by the three sites $i,j,k$. Let $\mathbf{r}_{i,j,k}$ be the initial positions of the three vertices, and $\mathbf{u}_{i,j,k}$ be the corresponding infinitesimal displacements. Denote by $\mathbf{u}_{(ij)}\equiv\mathbf{u}_j - \mathbf{u}_i$ and $\mathbf{r}_{(ij)}\equiv\mathbf{r}_j - \mathbf{r}_i$ the differences. We consider the infinitesimal change in the normal vector $\mathbf{n}_{(ijk)}$ of the panel bounded by vertices $i,j,k$. Set $\mathbf{n}^{(0)}_{(ijk)}$ to be the initial normal vector, given by
\begin{align}
    \mathbf{n}^{(0)}_{(ijk)}=\mathbf{r}_{(kj)}\times\mathbf{r}_{(ij)},
\end{align}
which is not necessarily normalized. The new normal vector induced by $\mathbf{u}_{i,j,k}$ is given by
\begin{multline}
\mathbf{n}^{(0)}_{(ijk)}+\delta\mathbf{n}_{(ijk)}=(\mathbf{r}_{(kj)}+\mathbf{u}_{(kj)})\times(\mathbf{r}_{(ij)}+\mathbf{u}_{(ij)})
\\
=\mathbf{r}_{(kj)}\times\mathbf{r}_{(ij)}+\mathbf{u}_{(kj)}\times\mathbf{r}_{(ij)}+\mathbf{r}_{(kj)}\times\mathbf{u}_{(ij)}+\mathbf{u}_{(kj)}\times\mathbf{u}_{(ij)}
\end{multline}
where the last term can be dropped since it is of higher order. Hence, the infinitesimal change in the normal vector is given by
\begin{align}
\delta\mathbf{n}_{(ijk)}=\mathbf{u}_{(kj)}\times\mathbf{r}_{(ij)}+\mathbf{r}_{(kj)}\times\mathbf{u}_{(ij)}.
\label{eq:delta_n}
\end{align}

On the other hand, the unit normal vector is given by 
\begin{multline}
\hat{\mathbf{n}}_{(ijk)}=\frac{\mathbf{n}_{(ijk)}^{(0)}+\delta\mathbf{n}_{(ijk)}}{\sqrt{(\mathbf{n}_{(ijk)}^{(0)}+\delta\mathbf{n}_{(ijk)})\cdot(\mathbf{n}_{(ijk)}^{(0)}+\delta\mathbf{n}_{(ijk)})}}
\\
=\frac{\mathbf{n}_{(ijk)}^{(0)}+\delta\mathbf{n}_{(ijk)}}
{\sqrt{\mathbf{n}^{(0)}_{(ijk)}\cdot\mathbf{n}^{(0)}_{(ijk)}+2\mathbf{n}^{(0)}_{(ijk)}\cdot\delta\mathbf{n}_{(ijk)}+\delta\mathbf{n}_{(ijk)}\cdot\delta\mathbf{n}_{(ijk)}}}
\end{multline}
where higher-order terms are dropped.

Again, we drop the last term in the denominator because it is of higher order. Then we can perform a Taylor expansion to get
\begin{align}
\hat{\mathbf{n}}_{(ijk)}=\frac{\mathbf{n}_{(ijk)}^{(0)}+\delta\mathbf{n}_{(ijk)}}{|\mathbf{n}_{(ijk)}^{(0)}|}\rb{1-\frac{\mathbf{n}_{(ijk)}^{(0)}\cdot\delta\mathbf{n}_{(ijk)}}{\mathbf{n}_{(ijk)}^{(0)}\cdot\mathbf{n}_{(ijk)}^{(0)}}}.\label{eq:nhat}
\end{align}

 We do the same thing to obtain a similar formula for the unit normal vector $\hat{\mathbf{n}}_{(jkl)}$ of the other panel. The new dihedral angle $\theta=\theta^{(0)}+\delta\theta$ between the two panels should satisfy
\begin{align}
\hat{\mathbf{n}}_{(ijk)}\cdot\hat{\mathbf{n}}_{(jkl)}=\cos(\theta^{(0)}+\delta\theta)\sim\cos\theta^{(0)}-\sin\theta^{(0)}\delta\theta
\end{align}

 Using the initial condition where $\theta^{(0)}$ satisfies $\cos\theta^{(0)}=\hat{\mathbf{n}}_{(ijk)}^{(0)}\cdot\hat{\mathbf{n}}_{(jkl)}^{(0)}$, we develop a linear mapping from $\delta\mathbf{n}_{(ijk)}$ and $\delta\mathbf{n}_{(jkl)}$ to $\delta\theta$ via
 \begin{multline}
    -\sin\theta^{(0)}\delta\theta=\hat{\mathbf{n}}_{(ijk)}^{(0)}\cdot\frac{\delta\mathbf{n}_{(jkl)}}{|\mathbf{n}_{(jkl)}^{(0)}|}+\hat{\mathbf{n}}_{(jkl)}^{(0)}\cdot\frac{\delta\mathbf{n}_{(ijk)}}{|\mathbf{n}_{(ijk)}^{(0)}|}\\-\hat{\mathbf{n}}_{(ijk)}^{(0)}\cdot\hat{\mathbf{n}}_{(jkl)}^{(0)}\rb{\hat{\mathbf{n}}_{(ijk)}^{(0)}\cdot\frac{\delta\mathbf{n}_{(ijk)}}{|\mathbf{n}_{(ijk)}^{(0)}|}+\hat{\mathbf{n}}_{(jkl)}^{(0)}\cdot\frac{\delta\mathbf{n}_{(jkl)}}{|\mathbf{n}_{(jkl)}^{(0)}|}.
    }\label{eq:dtheta}
 \end{multline}

Substitute the $\delta\mathbf{n}_{(ijk)}$ and $\delta\mathbf{n}_{(jkl)}$ as in \myeq{delta_n} into\myeq{dtheta}, we obtain a linear mapping from $\mathbf{u}_{i,j,k,l}$ to the infinitesimal change in dihedral angle $\delta\theta$. 

Each such linear map from the displacements to a change in a dihedral angle generates a row of $\mathbf{S}$, just as the equation for edge extensions~\cite{doi:10.1073/pnas.2005089117,mao2018maxwell, Rocklin2020} generates a row of the rigidity matrix. As in that case, $\mathbf{S}$ is a function of $\mathbf{z}$ because the displacement vectors $\mathbf{u}$ sometimes lie in different unit cells from the adjoining edge.

\section{Derivation of scaling laws of energy functional and stiffness}

For a topological origami sheet with lines of zero modes present in the Brillouin zone, we want to derive a scaling law for the response $\mathbf{u}_\mathbf{q}$. We begin with  the force-balance equation
\begin{align}
\textbf{f}(\textbf{q})=\nabla_{\mathbf{u}}E(\mathbf{q})=k_f\textbf{S}^T(-\textbf{q})\textbf{S}(\textbf{q})\textbf{u}_{\textbf{q}}+k_s\textbf{R}^T(-\textbf{q})\textbf{R}(\textbf{q})\textbf{u}_{\textbf{q}}
\label{eq:force_balance}
\end{align}
where $\mathbf{f}(\mathbf{q})$ is the Fourier transform of the external force applied to the system, and $k_s,k_f$ represent respectively the stiffness of stretching and folding. In the main text we use two diagonal matrices $\textbf{K}_\textbf{s}, \textbf{K}_\textbf{f}$ because the exact values of the diagonal entries vary from face to face, but they should be of the same order of magnitude and should not change the result as far as a scaling law is concerned. Hence, we simplify the situation by setting $\textbf{K}_\textbf{s}=k_s\mathbb{I},\textbf{K}_\textbf{f}=k_f\mathbb{I}$. 

As discussed in the main text, for topological origami there are curved lines of wavevectors in the Brillouin zone for which there are \emph{two} modes that do not stretch any panels and hence lie in the nullspace of $\mathbf{R}^T(\mathbf{-q})\mathbf{R}(\mathbf{q})$. However, for sheets of finite size, the actual modes present will not in general lie on these curved lines. Instead, they can differ by a small wavevector $\delta\mathbf{q}$, hence having eigenvalues on the order of $ |\delta\mathbf{q}|^2$.

We choose to examine the force-balance equation in the basis such that $\mathbf{R}^T(-\mathbf{q})\mathbf{R}(\mathbf{q})$ is diagonal. Because of the two zero modes, this matrix, which we denote $\tilde{\mathbf{D}}_s$ consequently has diagonal entries of the form $(b_1 |\delta \mathbf{q}|^2, b_2 |\delta \mathbf{q}|^2,b_3,b_4,\ldots)$, where $b_i$ are dimensionless constants on the order of one. In contrast, $\mathbf{S}^T(-\mathbf{q})\mathbf{S}(\mathbf{q})$ in this basis, which we denote $\tilde{\mathbf{D}}_b$, is not diagonal and has entries on the order of one, scaled by a factor of $A^{-1}$, where $A$ is the average panel area, as defined in the main text.

In the new basis formed by eigenvectors of $\mathbf{R}^T(-\mathbf{q})\mathbf{R}(\mathbf{q})$,  the force balance equation\myeq{force_balance} becomes
\begin{align}
\tilde{\mathbf{f}}=((k_f/A)\tilde{\mathbf{D}}_f+k_s\tilde{\mathbf{D}}_s)\tilde{\mathbf{u}}\equiv \mathbf{M}\Tilde{\mathbf{u}}
\label{eq:force_balance2}
\end{align} 
in which all quantities, except the two stiffness constants $k_f$ and $k_s$, should depend on the specific wavevector $\mathbf{q}$, but we omit it for simplicity in notation. Also, we bring the area factor $1/A$ in front of $\tilde{\mathbf{D}}_f$ so that $\tilde{\mathbf{D}}_f$ now consists of dimensionless numbers of order one.

Solving\myeq{force_balance2} exactly still involves inverting a $12\times12$ matrix and seems intractable at first sight. However, we can apply Cramer's rule to find qualitatively how the solution scales. Cramer's rule states that 
\begin{align}
    \tilde{u}_i=\frac{\det(\mathbf{M}_i)}{\det(\mathbf{M})}
\end{align}
where $\mathbf{M}_i$ is obtained by replacing the $i$-th column of $\mathbf{M}$ by $\Tilde{f}$. In the case $k_s\gg k_f/A$, the stretching part $k_s\Tilde{\mathbf{D}}_s$ dominates in $\mathbf{M}$. Notice that $\tilde{\mathbf{D}}_s$ is by definition a diagonal matrix. It follows that the diagonal terms should dominate in the expansion of $\det(\mathbf{M})$ and $\det(\mathbf{M}_i)$. Denote by $a_i$ the $i$-th diagonal element of $\tilde{\mathbf{D}}_f$ and by $m_i$ that of $\mathbf{M}$ we have
\begin{align}
    \tilde{u}_i\approx \frac{m_1m_2\cdots\Tilde{f}_i\cdots m_{12}}{m_1m_2\cdots m_{12}}
    \label{eq:cramer}
 \end{align}
where in the numerator the $i$-th term in the product is replaced by $\tilde{f}_i$. By definition, we have $m_i=a_ik_f/A+b_ik_s|\delta\mathbf{q}|^2$ for $i=1,2$, and $m_i=a_ik_f/A+b_ik_s$ for $i>2$. Putting those values back to\myeq{cramer}, it is not hard to see that, given $|\delta\mathbf{q}|^2\to 0$, we have
\begin{align}
    \tilde{u}_{1},\tilde{u}_2\gg\tilde{u}_3,\tilde{u}_4,\cdots,\tilde{u}_{12}
\end{align}
and the first two components should scale as
\begin{align}
    \Tilde{u}_i\approx\frac{\tilde{f}_i}{a_ik_f/A+b_ik_s|\delta\mathbf{q}|^2}, \ \ \ \ i=1,2
    \label{eq:scaling}
\end{align}
which is the same scaling law as we show in the main text, despite some minor changes in the notation.

Note that in the case the expression\myeq{scaling} diverges as both $|\delta \mathbf{q}|$ and $k_f$ approach zero, indicating the zero mode is reached and bending becomes free so that the ``zero mode'' actually costs zero energy. 

The energy corresponding to this mode at wavevector $\mathbf{q}$ is thus approximated by
\begin{align}
\label{eq:energy_q}
    E(\mathbf{q})\approx \sum_{i=1,2}  \frac{\tilde{f}_i^2}{a_i (k_f/A)+b_i k_s|\delta\mathbf{q}|^2}.
\end{align}

In the infinite-size system limit, the total energy is an integral of $E(\mathbf{q})$ in the Brillouin zone, and its leading order is obtained by substituting the integrand by our scaling law\myeq{energy_q}. So far as a qualitative scaling is concerned, we drop the subscripts of the dimensionless constants $a_i,b_i,f_i$ and only integrate one term in\myeq{energy_q} to get
\begin{align}
\label{eq:E_tot}
    E_\mathrm{tot}=
    \int_{BZ}E(\mathbf{q})d^2\textbf{q}\approx\int_{BZ} \frac{f^2}{a(k_f/A)+bk_s|\delta\mathbf{q}|^2} d^2 \textbf{q}.
\end{align}
Here we emphasize the difference between $d^2 \mathbf{q}$ and $|\delta\textbf{q}|^2$ to prevent possible confusion in the notations: the former is the 2-D infinitesimal area element $d^2 \mathbf{q}=dq_1dq_2$ in the Brillouin zone, while the latter is the square of the distance $|\delta\mathbf{q}|$ between that particular wavevector $\mathbf{q}$ and its nearest zero mode.
\begin{figure}[H]
    \centering\includegraphics[width=0.75\textwidth]{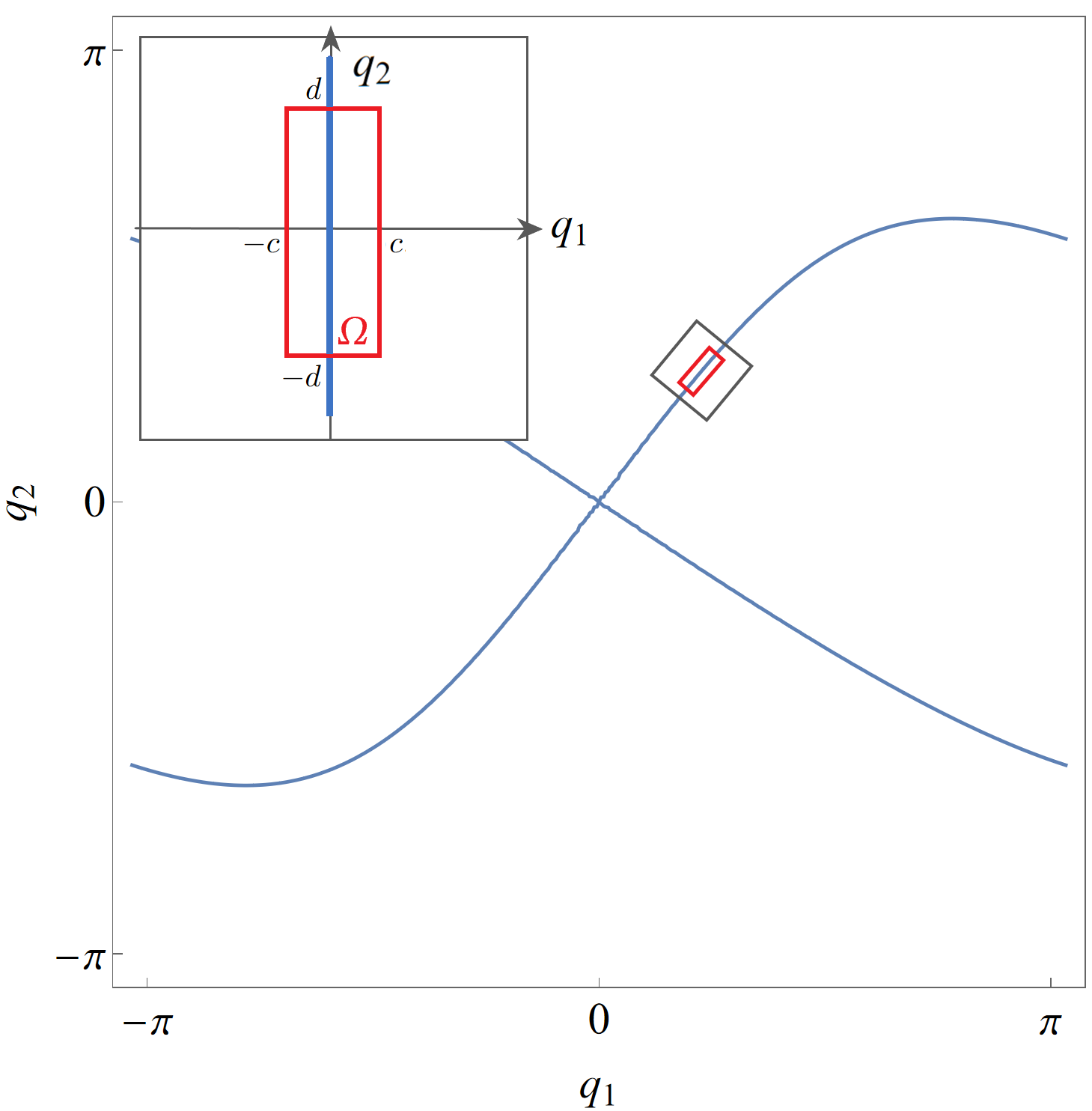}
    \caption{Approximation of $E_\mathrm{tot}$ by contribution near lines of zero modes. The majority of the energy comes from boxes like $\Omega$ that are close to lines of zero modes.}
    \label{energy_integral_approximation.png}
\end{figure}
To approximate the integral\myeq{E_tot}, we make two observations. First, since Lebesgue measure is both translational and rotational invariant, we may take the line of zero modes to be locally straight and align it with one of the axes $q_{1,2}=0$. Here we take $q_1=0$ for instance as shown in FIG. \ref{energy_integral_approximation.png}. In this case, the 2-D distance $|\delta\mathbf{q}|$ reduces to a 1-D distance $|\delta q_2|$, and the infinitesimal elements $d^2\mathbf{q}$ can be identified with $d^2(\delta\mathbf{q})$. The integral in a small rectangular region $A=[-c,c]\times[-d,d]$ near the line of zero modes becomes
\begin{multline}
\label{eq:integral_appro}
    \int_{BZ}E(\mathbf{q})d^2\textbf{q}\approx 
   \int_{\Omega}\frac{f^2}{a(k_f/A)+bk_s|\delta\mathbf{q}|^2} d^2 \textbf{q}\\
   =
   \int_{-d}^dd(\delta q_2)\int_{-c}^c\frac{f^2}{a(k_f/A)+bk_s(\delta{q_1})^2}d(\delta q_1).
\end{multline}

The second observation is that the integrand is highly localized near $\delta q_2=0$ given $k_f/A\ll k_s$. Therefore, at the cost of introducing only a minor error, we may extend the bound in the second integral from a finite interval $[-c,c]$ to infinity, in which case equation\myeq{integral_appro} is easily evaluted to be
\begin{align}
\label{eq:integral_appro2}
2d\int_{-\infty}^\infty\frac{1}{(k_f/A)+k_s(\delta{q_1})^2}d(\delta q_1)=
\frac{2\pi d}{\sqrt{abk_sk_f/A}}.\end{align}
The majority of energy contribution comes from boxes like $A$ along lines of zero mode, and every box region follows the same scaling as in\myeq{integral_appro2}. Hence, we conclude that the total energy should follow the power law ${(k_sk_f/A)}^{-1/2}$. The scaling law of the transitional case can be deduced in a similar manner.

\section{Simulations in MERLIN}

We run simulations in MERLIN~\cite{liu2018highly} using the ``displacement" and ``manual" mode. The ``ModelType" we use is ``N4B5". The ``MaterCalib" is set ``manual". The ``BarCM" uses a hyper elastic model ``Ogden" with the stiffness parameter taken to be $10^{-6}$. The ``Abar" is set $0.2$. The ``Kf" parameter is set $3\times10^{-4}$ in our simulations. The ``Kb" parameter does not affect the simulation results because all panels are fully triangulated in the processes. 

The three sheets we choose for numeric simulations have the following parameters. The conventional Miura sheet has $\alpha=\pi/4$ and the four edge lengths are all set $1$. The dihedral angle between the face spanned by $\mathbf{r}_1$ and $\mathbf{r}_2$ and the face spanned by $\mathbf{r}_2$ and $\mathbf{r}_3$ is set $4.78319$. In the Cartesian coordinate system, the two lattice vectors are $$l_1=(1.12647, 0, 0),$$$$l_2=(0,1.71151,0).$$ The four edge vectors in a unit cell have the following coordinates, respectively, 
$$\mathbf{r}_1=(0.563236, -0.826296, 0),$$
$$\mathbf{r}_2=(0,0.855754, -0.517382),$$
$$\mathbf{r}_3=(-0.563236, -0.826296, 0),$$
$$\mathbf{r}_4=(0,-0.855754, -0.517382).$$
The topological Eggbox sheet has $\alpha=107\pi/180$ and edge lengths $1$. The dihedral angle between the face spanned by $\mathbf{r}_1$ and $\mathbf{r}_2$ and the face spanned by $\mathbf{r}_2$ and $\mathbf{r}_3$ is set $4.28319$. In the Cartesian coordinate system, the two lattice vectors are $$l_1=(1.74065, 0, 0),$$$$l_2=(0,1.60941,0).$$ The four edge vectors in a unit cell have the following coordinates, respectively, 
$$\mathbf{r}_1=(0.870326, 0, 0.492475),$$
$$\mathbf{r}_2=(0, 0.804703, -0.593678),$$
$$\mathbf{r}_3=(-0.870326, 0, 0.492475),$$
$$\mathbf{r}_4=(0, -0.804703, -0.593678).$$
The transitional Morph sheet has $\alpha=1.85515$ and $\beta=1.45417$ and edge lengths $0.960397, 0.999085, 0.960397, 0.999085$. The dihedral angle between the face spanned by $\mathbf{r}_1$ and $\mathbf{r}_2$ and the face spanned by $\mathbf{r}_2$ and $\mathbf{r}_3$ is set $4.3013$. In the Cartesian coordinate system, the two lattice vectors are $$l_1=(1.83686, 0, 0),$$$$l_2=(0,1.68059,0).$$ The four edge vectors in a unit cell have the following coordinates, respectively, 
$$\mathbf{r}_1=(0.918432, -0.236166, 0.15189),$$
$$\mathbf{r}_2=(0, 0.840297, -0.540437),$$
$$\mathbf{r}_3=(-0.918432, -0.236166, 0.15189),$$
$$\mathbf{r}_4=(0, -0.840297, -0.540437).$$

\section{Second-order isometries}

In this appendix, we discuss an apparent connection between modes that are isometric to leading order in the wavevector and modes that are isometric to second order in the deformation. Our analysis follows from that of Ref.~\cite{doi:10.1073/pnas.2005089117}, which determines the condition for first order isometries to extend to second order in tessellations with periodic boundary conditions on their dihedral angles. However, as we show, the present case of parallelogram-based origami is distinct from the general case due to the existence of the in-plane mechanism. We rely on numerical rather than analytical computation, and therefore adopt the truss model that triangulates the panels to account for panel bending.

First, we review the isometries of a parallelogram-based origami sheet under periodic boundary conditions. In the truss model, there are eight real creases and four virtual creases. Hence, our isometries are a twelve-dimensional vector where we use $\varphi$ to denote folding on the real creases and $\delta$ to denote folding on the virtual creases. We choose our triangulation so that the virtual creases all emanate from the same vertex (i.e., there are two four-coordinated vertices and two eight-coordinated vertices). Given this triangulation, we construct the compatibility matrix (i.e., the equilibrium matrix of a spring-mass system $\mathbf{Q} = \mathbf{R}^T$ derived above) that maps the folding angles to constraints on the vertices. The isometries span the nullspace of this compatibility matrix, which turns out to be three dimensional. Since the mechanism does not bend any of the faces, its representation has $\delta=0$ for every virtual crease. We isolate this mode via orthogonalization from a generate mode in the nullspace of the compatibility matrix and refer to it as a strain mode. Simultaneously, there is a nonrigid isometry that is represented by $\varphi=0$ on every real crease for our choice of triangulation. We again isolate this mode via orthogonalization from the remaining two modes in the nullspace of the compatibility matrix. We use the remaining mode in the nullspace to form a basis for the three isometries. We refer to these two nonrigid isometries, respective to the order they were introduced, as the torsion and the curvature mode.

Second, we review the condition for linear isometries to extend to second order under periodic boundary conditions. The basis for the second order condition comes from the notion of lattice compatibility. In contrast to parallelogram-based origami, generic tessellations are generated by a combination of lattice vectors and lattice rotations that are required to satisfy the orientation and position compatibility conditions:

\begin{gather}
    \mathbf{S}_1 \mathbf{S}_2 = \mathbf{S}_2 \mathbf{S}_1, \\
    \boldsymbol{\ell}_1 + \mathbf{S}_1 \boldsymbol{\ell}_2 = \boldsymbol{\ell}_2 + \mathbf{S}_1 \boldsymbol{\ell}_1.
\end{gather}

\noindent
The physical significance of these conditions is that the periodic tessellation must adopt a quasi-cylindrical configuration. Any other geometry requires spatial heterogeneity. An isometry changes the lattice vectors ($\boldsymbol{\ell} \rightarrow \boldsymbol{\ell} + \boldsymbol{\Delta}$) and the lattice rotations ($\mathbf{S} \rightarrow (\mathbf{1} + \boldsymbol{L}) \mathbf{S}$, so that the expansion of these equations yields:

\begin{gather}
    \mathbf{L}_1 \mathbf{S}_1 \mathbf{S}_2 + \mathbf{S}_1 \mathbf{L}_2 \mathbf{S}_2 = \mathbf{L}_2 \mathbf{S}_2 \mathbf{S}_1 + \mathbf{S}_2 \mathbf{L}_1 \mathbf{S}_1, \\
    \boldsymbol{\Delta}_1 + \mathbf{L}_1 \mathbf{S}_1 \boldsymbol{\ell}_2 + \mathbf{S}_1 \boldsymbol{\Delta}_2 = \boldsymbol{\Delta}_2 + \mathbf{L}_2 \mathbf{S}_2 \boldsymbol{\ell}_1 + \mathbf{S}_2 \boldsymbol{\Delta}_1.
\end{gather}

\noindent
For the special case of quasi-planar tessellations (e.g., parallelogram-based origami), $\mathbf{S}_1 = \mathbf{S}_2 = \mathbf{1}$ in the ground state and therefore the orientation compatibility condition is trivially satisfied at first order. Consequently, the first order condition is insufficient to guarantee that the tessellation stays quasi-cylindrical and the following second order condition is required:

\begin{equation}
    \mathbf{L}_1 \mathbf{L}_2 = \mathbf{L}_2 \mathbf{L}_1
\end{equation}

\noindent
Generally, there are linear three isometries for these tessellations and this condition is a constraint on the amplitudes of a generic superposition of modes $\varphi = \sum \lambda_i \varphi_i$. Since the constraint is second order, solutions to $\lambda_3$ are given by the quadratic formula in terms of $\lambda_{1,2}$ as:

\begin{gather}
    c_1 \lambda_1^2 + c_2 \lambda_2^2 + c_3 \lambda_3^2 + c_4 \lambda_1 \lambda_2 + c_5 \lambda_1 \lambda_3 + c_6 \lambda_2 \lambda_3 = 0, \\
    \implies \lambda_3^\pm = -\frac{c_5 \lambda_1 + c_6 \lambda_2}{2 c_3} \pm \frac{1}{2 c_3}\sqrt{(c_5 \lambda_1 + c_6 \lambda_2)^2 - 4 c_3 (c_1 \lambda_1^2 + \lambda_2 (c_4 \lambda_1 + c_2 \lambda_2)},
\end{gather}

\noindent
where the coefficients $c_i$ depend on the geometry of the crease pattern. The two roots correspond to upwards and downwards bending cylindrical deformations. However, it is necessary that the solution be real and, as we show below, certain crease patterns can prevent the surd $(c_5 \lambda_1 + c_6 \lambda_2)^2 - 4 c_3 (c_1 \lambda_1^2 + \lambda_2 (c_4 \lambda_1 + c_2 \lambda_2)$ from being positive so that no linear combinations of the isometries are compatibility at second order. Once an admissible second order isometry is determined, the symmetry axis $\hat{S}$ of the induced quasi-cylindrical configuration is invariant under $\mathbf{L}_{1,2}$.

Third, we numerically compute$\mathbf{L}_{1,2}$ for the strain, torsion, and curvature modes in a generic parallelogram-based origami sheet with four faces. Naturally, these quantities vanish for the strain mode so that the linear combination has nontrivial contributions from the torsion and curvature modes exclusively. We find that when the bend mode generates synclastic curvature (resembling a dome) to first order there is a solution to the second order condition. However, when the bend mode generates anticlastic curvature (resembling a saddle) to first order, there is no solution to the second order condition. For example, using the miura geometry presented in the appendix above, we find:

\begin{gather}
    \mathbf{L}_1^s = \mathbf{0}, \quad \mathbf{L}_2^s = \mathbf{0}, \\
    \mathbf{L}_1^t \approx 0.398 \hat{L}_x, \quad \mathbf{L}_2^t \approx -0.605 \hat{L}_y, \\
    \mathbf{L}_1^c \approx -0.409 \hat{L}_y, \quad \mathbf{L}_2^c \approx -0.289 \hat{L}_x,
\end{gather}

\noindent
where $\hat{L}_{x,y}$ denote the generators of rotation about the $x,y$ axes and the superscripts $s,t,c$ denote the strain, torsion, and curvature modes respectively. The only solutions to the second order condition require $\lambda_c / \lambda_t \approx \pm 1.43i$ indicating there are no real second order isometries for the crease pattern. In contrast, using the eggbox geometry presented in the appendix above, we find:

\begin{gather}
    \mathbf{L}_1^s = \mathbf{0}, \quad \mathbf{L}_2^s = \mathbf{0}, \\
    \mathbf{L}_1^t \approx 0.732 \hat{L}_x, \quad \mathbf{L}_2^t \approx -0.676 \hat{L}_y, \\
    \mathbf{L}_1^c \approx -0.547 \hat{L}_y, \quad \mathbf{L}_2^c \approx0.861 \hat{L}_x.
\end{gather}

\noindent
Here, we find solutions to the second order condition given by $\lambda_c / \lambda_t \approx \pm 1.02$. Consequently, there are real second order isometries and we compute their symmetry axes as $\hat{S} \approx 0.793 \hat{x} + 0.608 \hat{y}$.

Fourth and finally, we conclude with the apparent relationship between modes that are isometric to second order in the deformation and modes that are isometric to leading order in the wavevector. Based on our intuition from the above analysis and from the results of the main text that reveal the same dichotomy between miura and eggbox patterns, we suppose that in the long wavelength limit the topological isometries are spatially varying cylindrical modes. To show this, we take the induced cylinder axis from above and determine the cell indices $(n_1,n_2)$ that satisfy:

\begin{equation}
    n_1 \boldsymbol{\ell}_1 + n_2 \boldsymbol{\ell}_2 = \hat{S}.
\end{equation}

\noindent
While these take the non-integer values $n_1 \approx 0.456$ and $n_2 \approx 0.378$, they inform the role of the wavevector in the Brillouin zone via the inner product $(n_1, n_2) \cdot (q_x, q_y)$. We take these cell indices and rotate by $\pi/2$ to determine the orthogonal direction $(n_1^*, n_2^*) \approx (-0.378, 0.456)$ and compute the determinant of the compatibility matrix as a function of the magnitude of this vector. We find that the determinant vanishes to eighth order in the magnitude, whereas for a generic direction the determinant vanishes only to fourth order. Thus, we deduce this is the soft direction in the long wavelength limit.

\bibliographystyle{unsrt}
\bibliography{bibliography}